\documentclass[preprint2]{aastex}
\usepackage{graphicx}
\usepackage{subfigure}
\usepackage{natbib}
\usepackage{epstopdf}

\shorttitle{Grain Size Distribution}
\shortauthors{Pauly \& Garrod}

\begin{document}


\title{The Effects of Grain Size and Temperature Distributions on the Formation of Interstellar Ice Mantles}


\author{Tyler Pauly\altaffilmark{1} and Robin T. Garrod\altaffilmark{1,2}} \email{tap74@cornell.edu}


\altaffiltext{1}{Cornell Center for Astrophysics and Planetary Science, Cornell University, Ithaca, NY 14853-6801}
\altaffiltext{2}{Departments of Astronomy \& Chemistry, University of Virginia, Charlottesville, VA 22904}


\begin{abstract}
Computational models of interstellar gas-grain chemistry have historically adopted a single dust-grain size of 0.1 micron, assumed to be representative of the size distribution present in the interstellar medium. Here, we investigate the effects of a broad grain-size distribution on the chemistry on dust-grain surfaces and the subsequent build-up of molecular ices on the grains, using a three-phase gas-grain chemical model of a quiescent dark cloud. We include an explicit treatment of the grain temperatures, governed both by the visual extinction of the cloud and the size of each individual grain-size population. We find that the temperature difference plays a significant role in determining the total bulk ice composition across the grain-size distribution, while the effects of geometrical differences between size populations appear marginal. We also consider collapse from a diffuse to a dark cloud, allowing dust temperatures to fall. Under the initial diffuse conditions, small grains are too warm to promote grain-mantle build-up, with most ices forming on the mid-sized grains. As collapse proceeds, the more abundant, smallest grains cool and become the dominant ice carriers; the large population of small grains means that this ice is distributed across many grains, with perhaps no more than 40 monolayers of ice each (versus several hundred assuming a single grain size). This effect may be important for the subsequent processing and desorption of the ice during the hot-core phase of star-formation, exposing a significant proportion of the ice to the gas phase, increasing the importance of ice-surface chemistry and surface-gas interactions.
\end{abstract}




\section[]{Introduction}

The chemistry of both quiescent and star-forming regions of the interstellar medium (ISM) is influenced significantly by processes occurring on the surfaces of the dust grains that permeate the gas \citep[e.g.][]{Whittet89,Chiar95,Geppert06,Boogert08,Cuppen09}. Atoms and molecules may accrete onto the grain surfaces, depleting gas-phase abundances, while allowing molecular ice mantles to form on the grains, typically through the addition of atomic hydrogen to atoms and simple molecules such as CO. The products of surface chemical reactions may be returned to the gas-phase by both thermal and non-thermal desorption, while the composition of the bulk ices may be further altered by processing associated with star formation. However, observations of infrared line absorption toward dark clouds and star-forming regions indicate that, while specific abundances may vary between individual sources, the major solid-phase repositiories of oxygen, carbon and nitrogen are a small group of simple, stable molecules: H$_2$O, CO, CO$_2$, CH$_4$, CH$_3$OH, H$_2$CO and NH$_3$ \citep[see][]{Gibb00,Whittet07,Oberg08}.

Various computational models of the combined gas-phase/grain-surface chemical system exist and are well described in the literature \citep[e.g.][]{Hasegawa92,Aikawa08,Garrod08,Wakelam10,Vasyunin09}. Common to almost all such models is the adoption of a single, representative dust-grain radius, typically 0.1 $\mu$m, with the associated number of surface binding sites on the order of 1 million. In fact, a size distribution of interstellar dust grains was defined based on observational data by \citet{Mathis77}, with the simple relationship $dn \propto a^{-3.5} da$, where $a$ is the radius of a spherical grain and $0.005 \mu m < a < 1.00 \mu m$. More recent work by \citet{Weingartner01} has defined the distribution more precisely.

The use in astrochemical models of a single grain size has, so far, worked adequately well for two main reasons: firstly, in the majority of astrophysical applications, there is a net deposition of material from the gas-phase onto the grains. The key parameter affecting the gas-phase chemistry is not the individual sizes of grains in the distribution, but the total accreting surface area per unit volume of gas, which can be defined without requiring a size distribution.

The second reason concerns the grain-surface chemistry itself. Most gas-grain models employ surface reaction rates of the form
\begin{equation}
R_{\mathrm{AB}} = [k_{\mathrm{hop}}(A) + k_{\mathrm{hop}}(B)] \, n(A) \, n(B) / n_{s}
\end{equation}
where $R_{AB}$ is the rate of the reaction between species A and B (assuming no activation barrier is present), $k_{\mathrm{hop}}$ is the inverse lifetime of a particular species to hop from one binding site to an adjecent site, $n(A)$ and $n(B)$ are the abundances of species A and B, and $n_{s}$ is the number of binding sites on the surface, taken in the same units as $n(i)$. Grain-surface abundances are typically parameterized as a fraction with respect to total hydrogen, following the usual treatment for gas-phase species. As explained by \citet{Acharyya11}, a simple chemical system consisting of accretion, evaporation and surface reactions governed by equation (1) is unaffected by the choice of representative grain size, so long as the surface area available for accretion is held constant.

However, the fidelity of such an approach may break down for several reasons. Firstly, as described by \citet{Acharyya11}, the growth of an ice mantle on the grain surface must result in an increase in the accretion cross section of the grains. This may be treated by modeling the grain/ice radius explicitly, but the increase of radius with ice mantle is nevertheless dependent on the assumed underlying grain size (as well as its morphology), requiring an explicit grain size to be chosen.

Secondly, equation (1) is not universally applicable to all surface reactions at all times, due to stochastic effects \citep[see][]{Garrod08}. These effects become important under conditions where both reactants in equation (1) attain surface populations close to or less than one per grain. Such conditions can become especially prevalent for small grains. Any method used to treat such effects explicitly, including the so-called modified-rate method, necessarily considers the species populations per grain, which again breaks the independence of the system from the choice of grain size.

Perhaps most importantly, the temperatures of the dust grains are strongly dependent on their sizes. A simple analysis, based on the assumption that the dust absorption efficiency for relevant wavelengths is of order unity \citep{Krugel03}, indicates that grain temperatures are proportional to $a^{-1/6}$, where $a$ is the dust-grain radius. As found by \citet{Garrod11}, while elements of the grain-surface chemistry are robust to small changes in temperature in the 8 -- 12 K range, others may vary drastically; the abundance and formation mechanism of surface CO$_2$ in particular was found to switch according to a threshold temperature of around 12 K, with values greater than this leading to highly efficient conversion of CO to $\mathrm{CO_2}$, while temperatures below this value produced much more moderate CO$_2$ formation. The inclusion of a distribution of grain sizes would allow grain-size populations to fall either above or below this threshold temperature, under static conditions. Furthermore, the collapse models of \citet{Garrod11} produced falling dust temperatures, as visual extinction increased, reducing the degree of grain heating by the external, interstellar radiation field. The consideration of a size distribution would allow different populations to reach different temperatures at each stage of collapse.

\citet{Acharyya11} used size distributions in gas-grain chemical models, finding no increase in agreement with observations of either gas or grain species. However, the dust temperatures were fixed, and held uniform across all grain sizes, and the models did not include a method to reproduce the stochastic behavior of the surface chemistry. 

In this study, we expand the approach of \citet{Garrod11}, to include a distribution of initial grain sizes, with the radii of each grain population increasing as ice mantles are formed. The grain radii influence the surface chemistry through the accretion rates of gas-phase species, through the determination of the absolute populations of reactants per grain, and in the calculation of grain temperatures, which affects the surface diffusion and desorption rates. We investigate the effects of the grain-size and temperature distributions on the formation of the major components of interstellar ices, under static conditions, as well as under the varying density, extinction and dust temperature conditions associated with collapse to form a dark-cloud core. 

The computational and modeling methods used are outlined in Section 2. Results are presented in Section 3, and discussed in Section 4. We highlight the main conclusions of this study in Section 5.

\section[]{Methods}

We use a version of the three-phase gas-grain model {\em MAGICKAL} \citep{Garrod13} to simulate the coupled gas-phase and grain-surface chemistry. The surface consists of one chemically-active monolayer, whose total allowed population varies with grain size (see Section 2.1).  Following \citet{Garrod11}, due to the low-temperature conditions used in the models, chemistry {\em within} the ice mantles themselves is switched off, as is bulk diffusion between ice surface and mantle; the ice mantle beneath the surface acts as an inert store of material. A net gain in surface atoms and molecules (due to reaction, desorption and accretion) results in a commensurate transfer of surface material into the mantle, with each chemical species being transferred in proportion to its abundance in the surface layer. A similar transfer occurs in the case of a net loss of surface atoms and molecules, with species transferred to the surface layer according to their abundances in the mantle.

The dust-grain chemical model is expanded to include all reactions and processes for an arbitrary number of grain-size populations, upon each of which an individual surface chemistry is traced. For this reason, the gas-phase and grain-surface chemical network is reduced somewhat from that used by \citet{Garrod13}, to allow faster numerical integration. This mainly involves the removal of carbon-chain species with greater than 5 atoms and of species and reactions related to the elements Si, Cl and P. The network includes a total of 475 gas-phase species, of which the 200 neutrals may also reside on the grain surfaces or within the ice mantles. The network includes photodissociation of surface species, as well as photo-desorption and reactive-desorption processes \citep[see][]{Garrod07,Garrod13}. Gas-phase photodissociation of H$_2$ and CO are treated following \citet{Lee96}, as in our previous models. The grain surface chemistry is treated using modified rate method ``C'', of \citet{Garrod08}.

\subsection{Grain-Size Distribution}

In the models presented here, we consider the grain population to be initially composed of a distribution of grain radii as defined by the classic description of \citet{Mathis77}, wherein 
\begin{equation} \label{eq:powerlaw}
dn = c \times a^{-3.5} da
\end{equation}
where $n$ is the population of the specific grain size $a$, and $c$ is a constant, with upper and lower limits to the distribution of 0.25 and 0.005 $\mu$m.  A value of $c=7.762\times 10^{-26}$ is taken from \citet{Draine84}, which assumes silicate grains.

Here, each grain is assumed to be spherical, having a cross-sectional area of $\sigma = \pi a^{2}$. The full size distribution is divided into $N_{\mathrm{b}}$ bins, equally-spaced across the range of $\log_{10} \sigma$. Since the pertinent quantity for the purposes of accretion and surface coverage is area, not radius, this binning method ensures optimal coverage of the full range of grain conditions. For each bin, $i$, the mean cross-sectional area of grains between its lower and upper limits, $<\sigma(i)>$, is calculated (using equation 2), and this value is used as the representative grain area for this bin, from which its radius -- and thus volume -- is also calculated.

\subsection{Grain Mantle Growth}

The new models also consider the time-dependent ice-mantle growth in the calculation of dust-grain radii. In line with our previous models, we assume an areal surface binding-site density of $1/A_{s}$ = $1.0 \times 10^{15}$ cm$^{-2}$, which provides the surface density of sites and surface species. In order to determine the physical depth of a surface monolayer, we simply assume $d_{\mathrm{ML}}=A_{s}^{1/2}$. 

As a dust grain gains material, its radius must grow. Thus, at each moment in model time, the radius of each dust-grain population is assessed, with each constituent atom or molecule of the ice mantle and surface layer adding a volume $A_{s}^{3/2}$ to the basic grain volume for each grain population. The effective radius, $a_{\mathrm{eff}}(i)$, of each grain is then re-calculated assuming sphericity. The number of binding sites on each grain-size population is thus also calculated at each moment, according to $N_{s}(i)=4 \pi a_{\mathrm{eff}}^{2}(i)/A_{s}$. This value is used in the calculation of surface chemical rates (Eq. 1). An effective cross-sectional area for each grain population is also calculated, $<\sigma_{\mathrm{eff}}(i)>$, which determines the rate of accretion of gas-phase material onto that grain population. 

\subsection{Grain Temperature}

Here, we present certain models that include a distribution of grain temperatures across the size distribution. In order to calculate the variation of temperature with effective radius, we use the simple approximation $T_{d}(i) \propto a_{\mathrm{eff}}^{-1/6}(i)$. Although this relationship is strictly only applicable in the case that the dust is fully efficient at absorbing relevant photons for heating and the dust temperature is smooth over time, it allows the general behavior of the grain-size distribution to be traced in a simple manner. A more accurate approach to grain temperatures at the smallest sizes may be incorporated into future models.

In the collapse models presented, the dust temperatures are determined according to the expression derived by \citet[equation 17]{Garrod11}. This expression refers to a canonical grain of 0.1 $\mu$m radius, whose temperature varies according to the heating--cooling balance of the grain at various visual extinctions. This temperature is also scaled using the above relationship to take account of the varying grain sizes of each grain population, as the ice mantles grow.

\section[]{Results}

To investigate the effects of the grain-size distribution, we present models that differ incrementally from three control models. Each such control model assumes a single grain size (i.e. $N_b=1$) with a temperature of 8, 10 or 12 K, a visual extinction of 10 and a gas density $\mathrm{n_H = 2 \times 10^4}$ $\mathrm{ cm^{-3}}$. All static dark cloud models are run at this density, and all models in this paper assume a fixed {\em gas} temperature of 10 K. The single-grain control models are labeled 1G\_T8, 1G\_T10, and 1G\_T12 respectively; see Table 1. Table 2 shows the initial and final effective grain sizes for these and other models. It should be noted that, due to the strong bias of Eq. (2) toward small grain sizes, the initial grain size of the single-grain models is $\sim$10 times smaller than the typically-adopted representative grain size in previous models.

The next set of models uses a distribution of 5 grain sizes ($N_b=5$), with a {\em uniform} dust-grain temperature of 8, 10 or 12 K across all grain sizes. These models are labeled 5G\_T8\_UNIF, 5G\_T10\_UNIF, and 5G\_T12\_UNIF.

The final set of dynamically static models assumes a dust-temperature distribution, with the initial temperature determined by grain radius, as described in Section 2. A systemic temperature is defined, of 8, 10 or 12 K, which refers to the temperature assigned to a grain with an effective size (i.e. grain + mantle) of the canonical 0.1 $\mu$m radius. The initial temperatures of each of the 5 grain populations are scaled to this temperature, and are then allowed to change according to the ice-mantle growth that modifies the effective grain radius in each population. The three models are labeled 5G\_T8\_DIST, 5G\_T10\_DIST, and 5G\_T12\_DIST.

We also present results from a free-fall collapse model, labeled 5G\_COLL, with an initial density $\mathrm{n_H = 3 \times 10^3}$ $\mathrm{cm^{-3}}$ and a final density $\mathrm{n_H = 2 \times 10^4}$ $\mathrm{cm^{-3}}$, as carried out by \citet{Garrod11}. The initial dust temperature is determined by the initial visual extinction, which is set to $\mathrm{A_V = 3}$ such that the model finishes at similar density and $\mathrm{A_V}$ to our dark cloud models. The associated increase in visual extinction results in a drop in the temperature of the canonical 0.1 $\mathrm{\mu m}$ grain from 14.72 K to 8.14 K.  In this model, the temperature of each grain population varies both with the increasing visual extinction, caused directly by collapse, and with the increase in grain-mantle size, which also acts to lower the grain temperature. Note that in any model where a temperature distribution is used which depends on grain size, the temperature may also vary as a result of grain-mantle growth.

\begin{deluxetable}{llrrr}
\tablecolumns{5}
\tablewidth{0pc}
\tablecaption{List of models and parameters used, where $\mathrm{N_G}$ is the number of grain sizes discretized from the grain size distribution (if used), $\mathrm{T_{D}[0.1\mu}$m] is the initial temperature of a dust grain with radius $= \mathrm{0.1\mu}$m, and $\mathrm{A_V}$ is the visual extinction. For the DIST and COLL models, initial and final values are given.}
\tablehead{
\colhead{Model} && \colhead{$\mathrm{N_G}$}  & \colhead{$\mathrm{T_{D}[0.1\mu}$m]}    & \colhead{$\mathrm{A_V}$}}
\startdata
 1G\_T8     && 	1 & 8 & 10  \\
 1G\_T10   && 	1 & 10 & 10  \\
 1G\_T12   && 	1 & 12 & 10  \\
 \hline
 5G\_T8\_UNIF  &&	5 & 8 & 10  \\
 5G\_T10\_UNIF && 	5 & 10 & 10  \\
 5G\_T12\_UNIF  && 5 & 12 & 10  \\
 \hline
 5G\_T8\_DIST & (i)  &	5 & 8  & 10  \\
 5G\_T8\_DIST & (f)  &	5 & 7.65  & 10  \\
 5G\_T10\_DIST & (i)   & 5 & 10 & 10  \\
 5G\_T10\_DIST & (f)   & 5 & 9.62 & 10  \\
 5G\_T12\_DIST & (i)   & 5 & 12 & 10  \\
 5G\_T12\_DIST & (f)   & 5 & 11.59 & 10  \\
 \hline
 5G\_COLL & (i)   & 5 & 14.72 & 3  \\
 5G\_COLL & (f)   & 5 & 8.14 & 10.63  \\
\enddata
\label{table:modelparams}
\end{deluxetable}

\begin{figure*}
     \begin{center}
        \subfigure[Model 1G\_10]{
            \label{fig:1GT10growth}
            \includegraphics[width=0.46\textwidth]{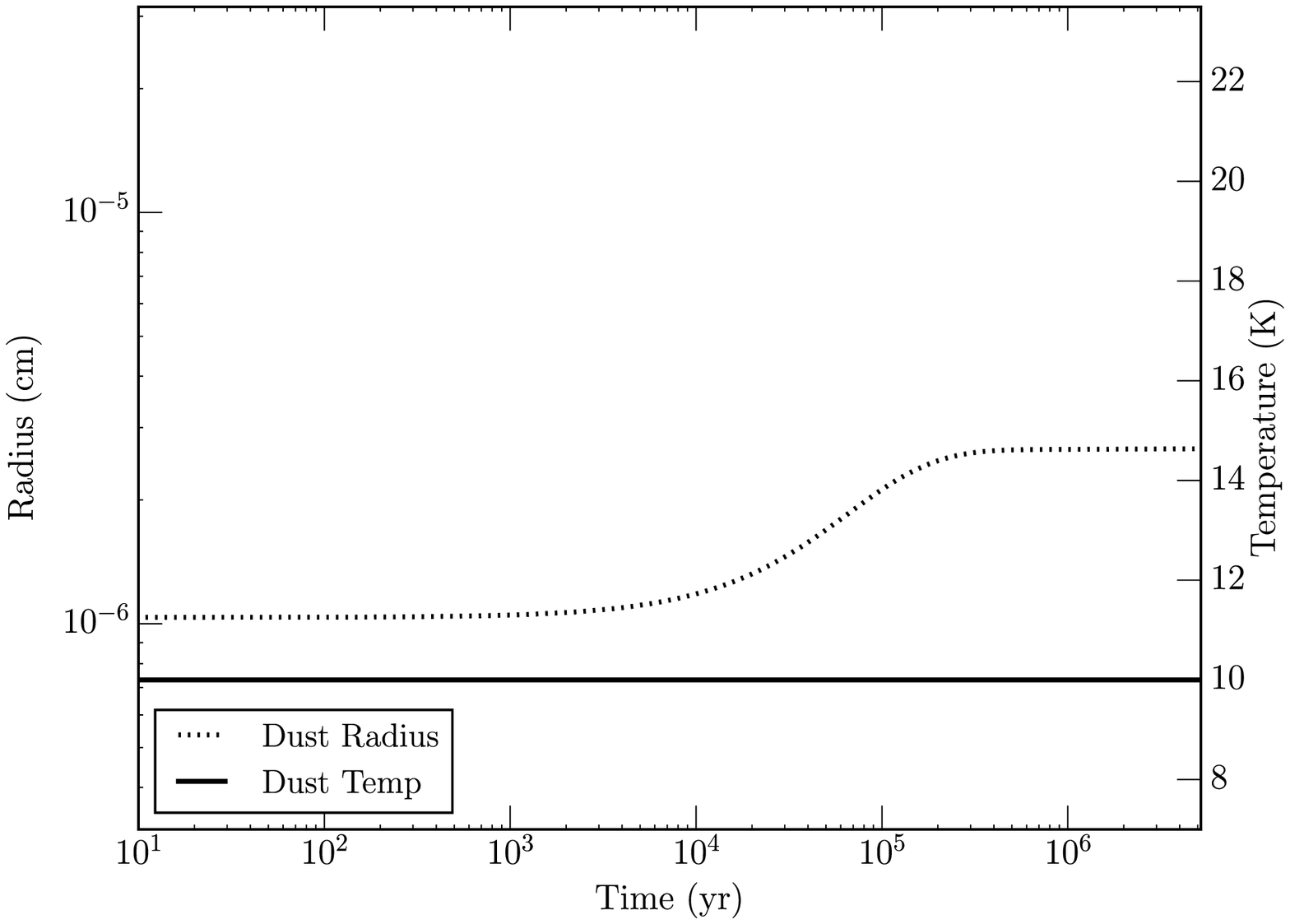}
          }
        \subfigure[Model 5G\_T10\_UNIF]{
           \label{fig:5G10Ugrowth}
           \includegraphics[width=0.46\textwidth]{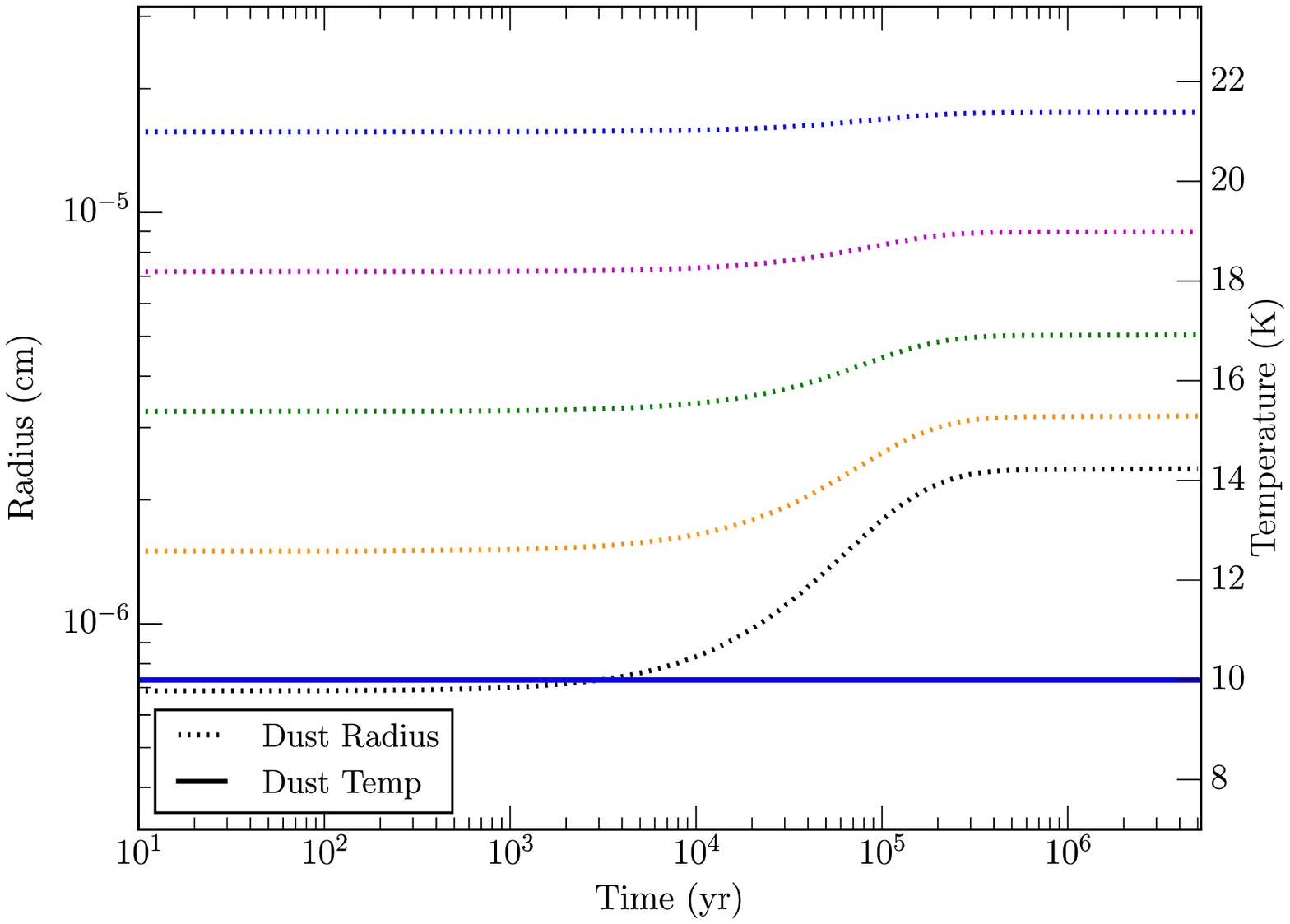}
        }\\ 
        \subfigure[Model 5G\_T10\_DIST]{
            \label{fig:5G10Dgrowth}
            \includegraphics[width=0.46\textwidth]{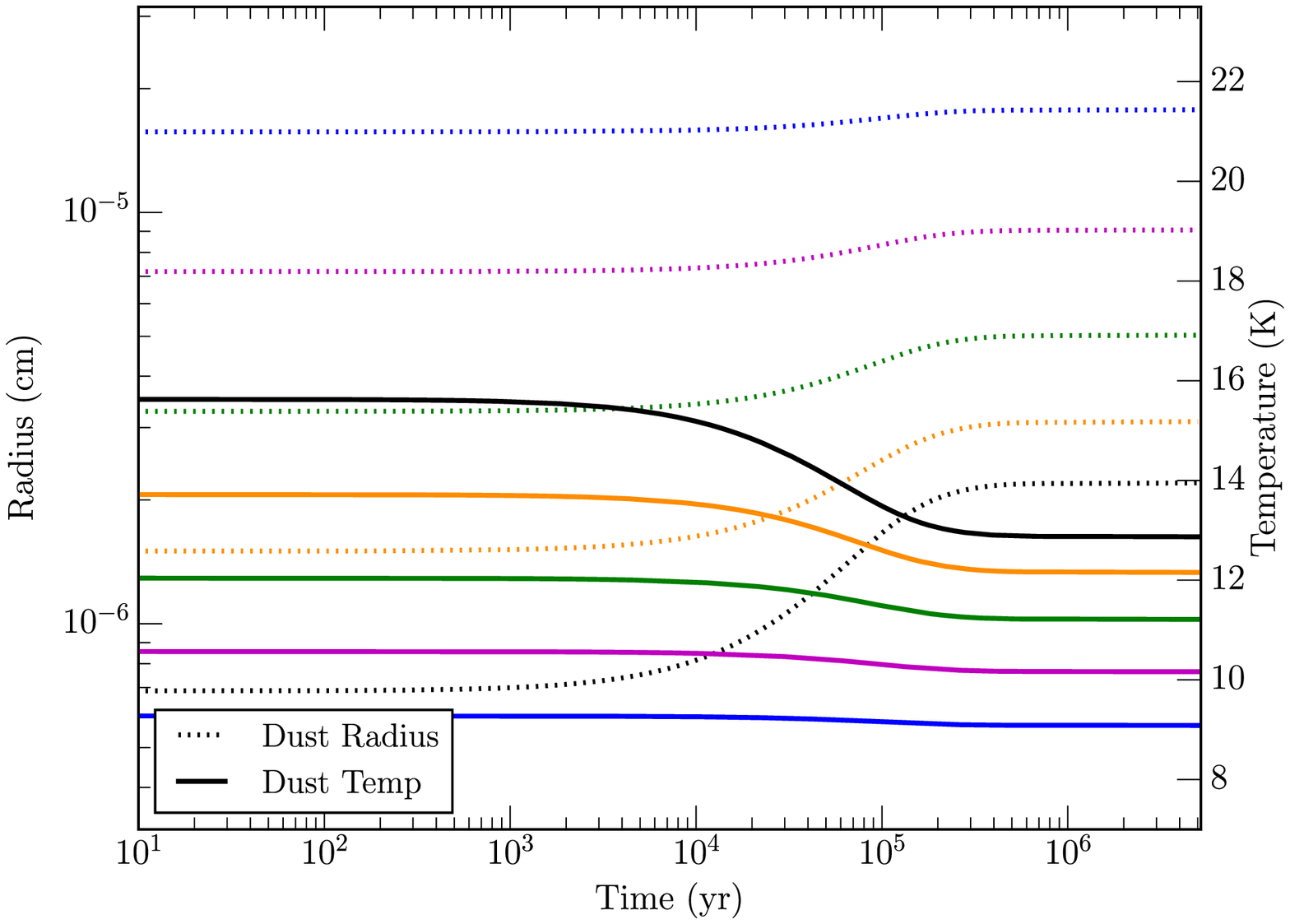}
        }
        \subfigure[Model 5G\_COLL]{
            \label{fig:2e4Collgrowth}
            \includegraphics[width=0.46\textwidth]{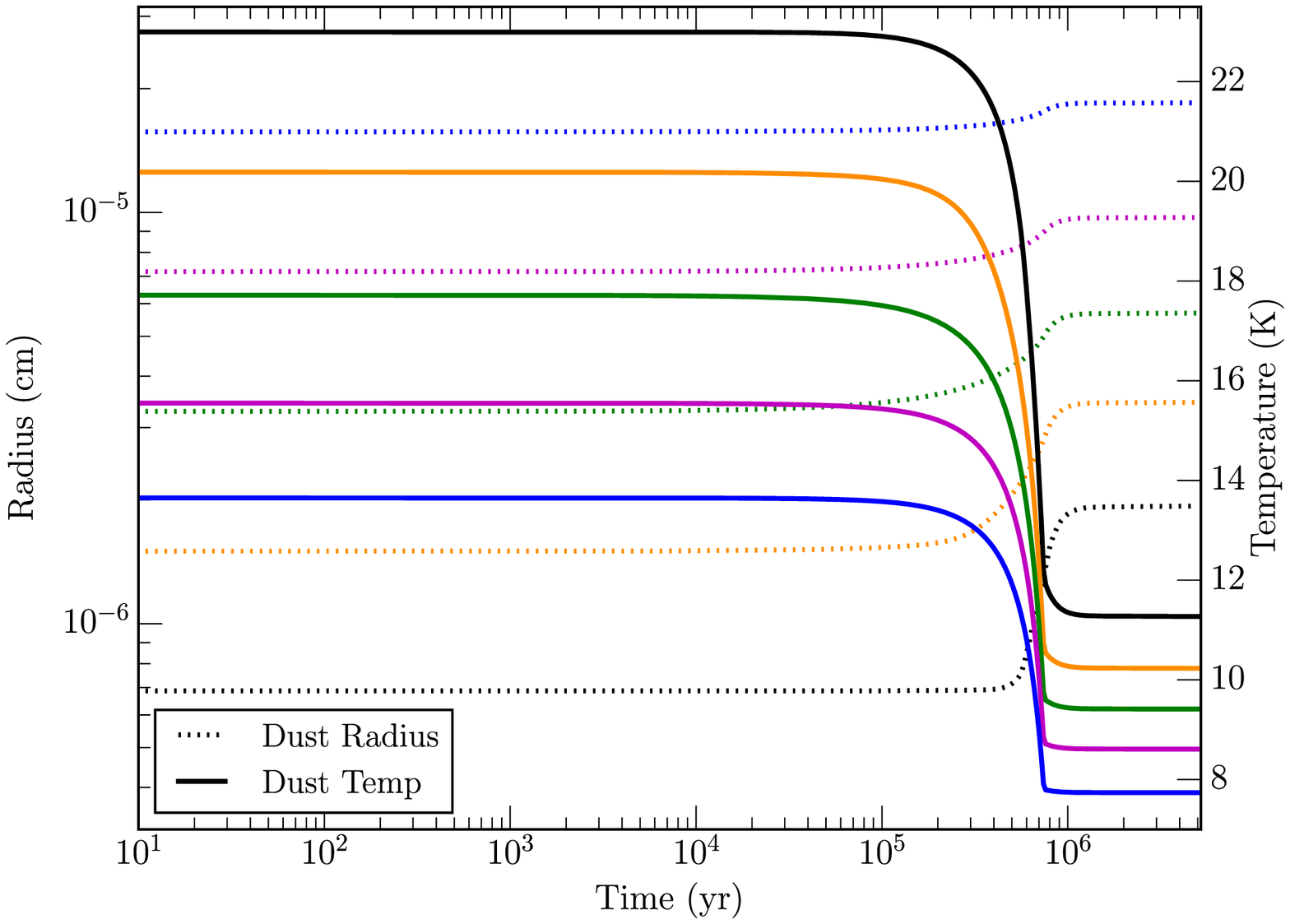}
        }
     \end{center}
     \caption{
        Grain size and temperature evolution for the four classes of models. Grain size is shown on the left side using the dashed lines, while the temperature is read off the right side using the solid lines. Colors correspond to grain sizes; temperature and grain size line colors correspond to the same grain population.
     }
     \label{fig:graingrowth}
\end{figure*}

\subsection[]{Mantle Growth and Temperature Evolution}

Figure~\ref{fig:graingrowth} shows the grain-size evolution for all of the various models.  Mantle growth occurs as gas species accrete onto grain surfaces and form ice mantles. Panel~\ref{fig:1GT10growth}, corresponding to the static, single-grain, $\mathrm{T_d=10}$ K model, shows significant growth beginning around $\mathrm{10^4}$ yr. This growth is seen at similar times in the five-grain model, shown in Panel~\ref{fig:5G10Dgrowth}. As growth is driven by accretion, mantle growth of a given grain size is determined primarily by the total cross-sectional area of that population.  Mantle growth is also modified by the grain temperature; if the temperature is sufficiently high that desorption of a common species is frequent, ice accumulation rates may be lower. This can be seen by comparing the total growth of the smallest grain for model 5G\_T10\_DIST (which uses a distribution of time-dependent dust temperature) to the growth of the same grain size in model 5G\_T10\_UNIF (in which all five grain sizes have temperature 10 K), as shown in Table~\ref{table:graingrowth}.

The smallest grains are shown to grow significantly, tripling their radius in most cases. Also, while an individual large grain has greater cross-sectional area, due to the lower population of large grains the dominant ice component is found on the smallest grains.

Figure~\ref{fig:graingrowth} also shows the $temperature$ evolution of the grains. For the single grain and uniform temperature models, we set a fixed temperature and no change occurs. The temperature distribution models show temperature evolution due to grain growth. Following the aforementioned power law, as grain radius increases the grain temperature drops. This temperature variation can be a strong determinant as to which chemical species are abundant, due to the temperature-sensitive nature of grain surface reactions.

\begin{deluxetable}{llrrrrr}
\tablecolumns{7}
\tablewidth{0pc}
\tablecaption{Initial grain size (i) and final grain size (f) in $\mathrm{\mu}$m for each grain size in all models used. Note that for a given number of grain sizes, all models start with the same initial radii.}
\tablehead{
\colhead{Model} & \colhead{}   & \colhead{$\mathrm{R_{gr1}}$}    & \colhead{$\mathrm{R_{gr2}}$} &
\colhead{$\mathrm{R_{gr3}}$}    & \colhead{$\mathrm{R_{gr4}}$}   & \colhead{$\mathrm{R_{gr5}}$}}
\startdata
1G\_**       & (i)  & 	0.0104 & & & & \\
\hline
1G\_T8       & (f)  &	0.0285 & & & &  \\
1G\_T10      & (f)  &	0.0266 & & & &  \\
1G\_T12      & (f)  &	0.0259 & & & &  \\
\hline
5G\_**** & (i) &  0.0069 & 0.015 & 0.033 & 0.072 & 0.157 \\
\hline
5G\_T8\_UNIF & (f)  &	0.0257 & 0.034 & 0.053 & 0.092 & 0.177 \\
5G\_T10\_UNIF& (f)  &	0.0238 & 0.032 & 0.050 & 0.090 & 0.175 \\
5G\_T12\_UNIF& (f)  &	0.0231 & 0.031 & 0.049 & 0.088 & 0.175  \\
\hline
5G\_T8\_DIST & (f)  &	0.0234 & 0.032 & 0.052 & 0.094 & 0.182\\
5G\_T10\_DIST& (f)  &	0.0220 & 0.031 & 0.050 & 0.091 & 0.178 \\
5G\_T12\_DIST& (f)  &	0.0224 & 0.031 & 0.048 & 0.086 & 0.176 \\
\hline
5G\_COLL     & (f)  &  0.0193 & 0.035 & 0.057 & 0.097 & 0.185 \\
\enddata
\label{table:graingrowth}
\end{deluxetable}

\subsection[]{Chemical Evolution - Single Grain}

\begin{figure*}
     \begin{center}
        \subfigure[Model 1G\_T8]{
           \label{fig:1G8growthcomp}
           \includegraphics[width=0.46\textwidth]{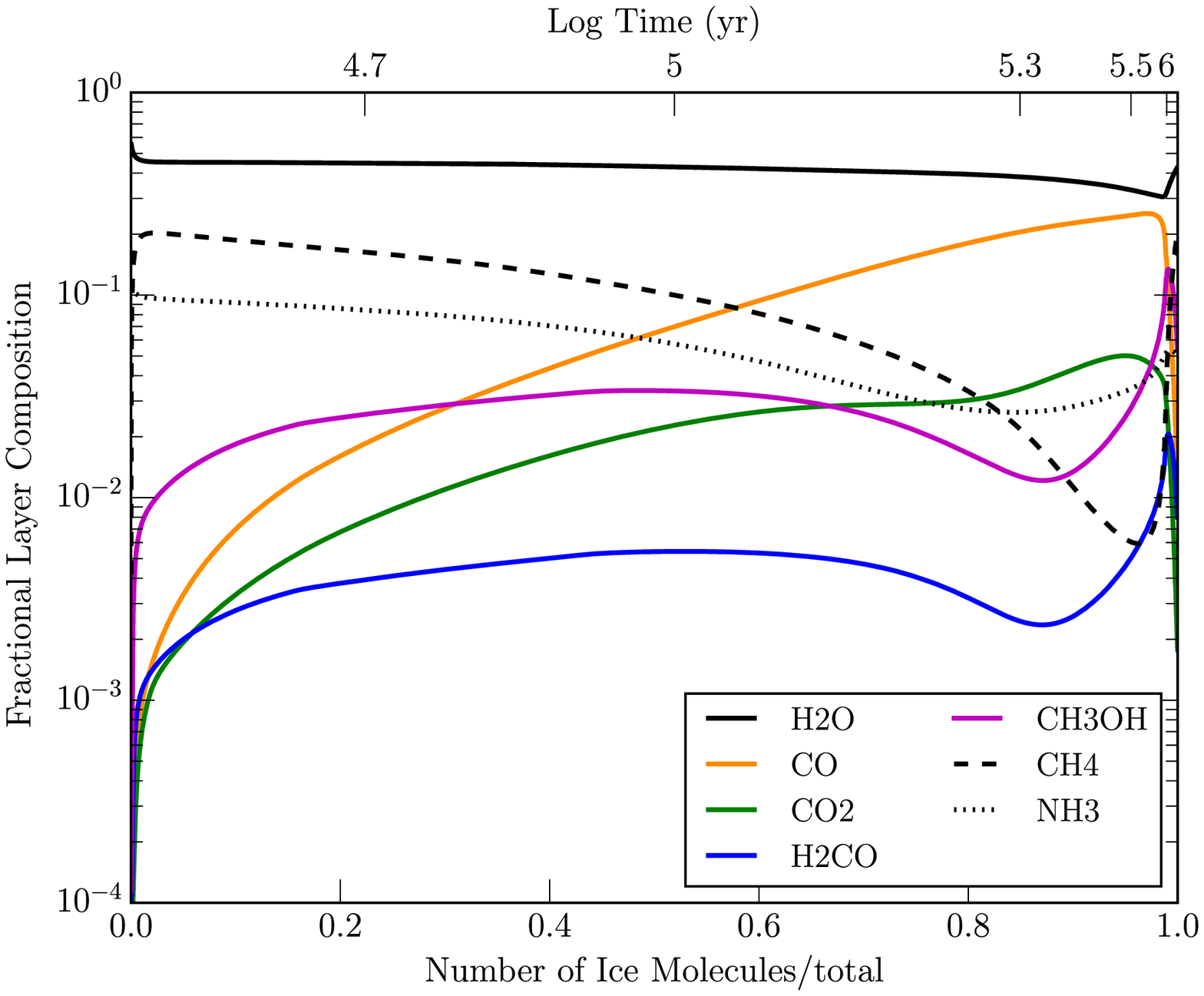}
        }
        \subfigure[Model 1G\_T10]{
            \label{fig:1GT10growthcomp}
            \includegraphics[width=0.46\textwidth]{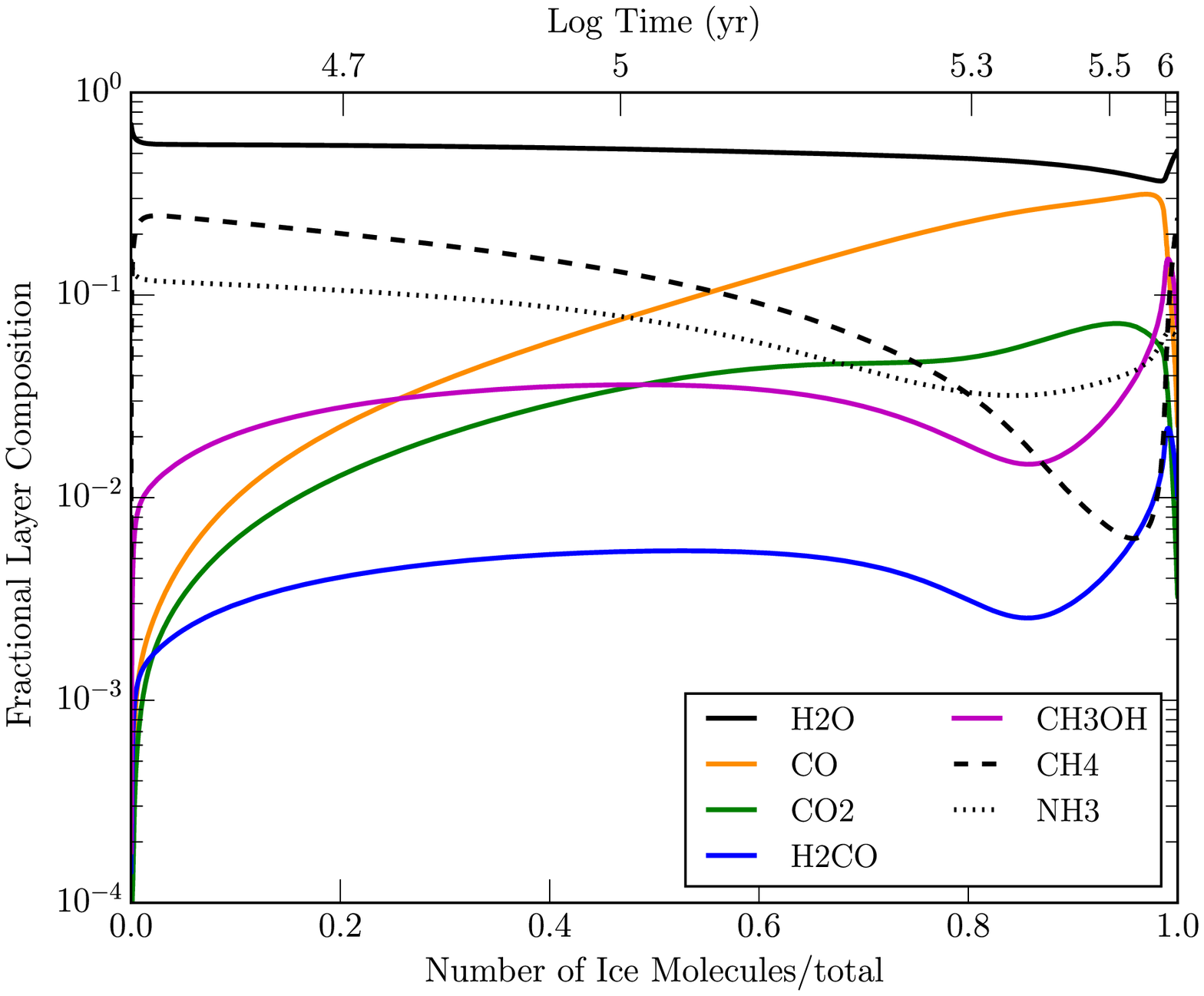}
        }\\ 
        \subfigure[Model 1G\_T12]{
            \label{fig:1G12growthcomp}
            \includegraphics[width=0.46\textwidth]{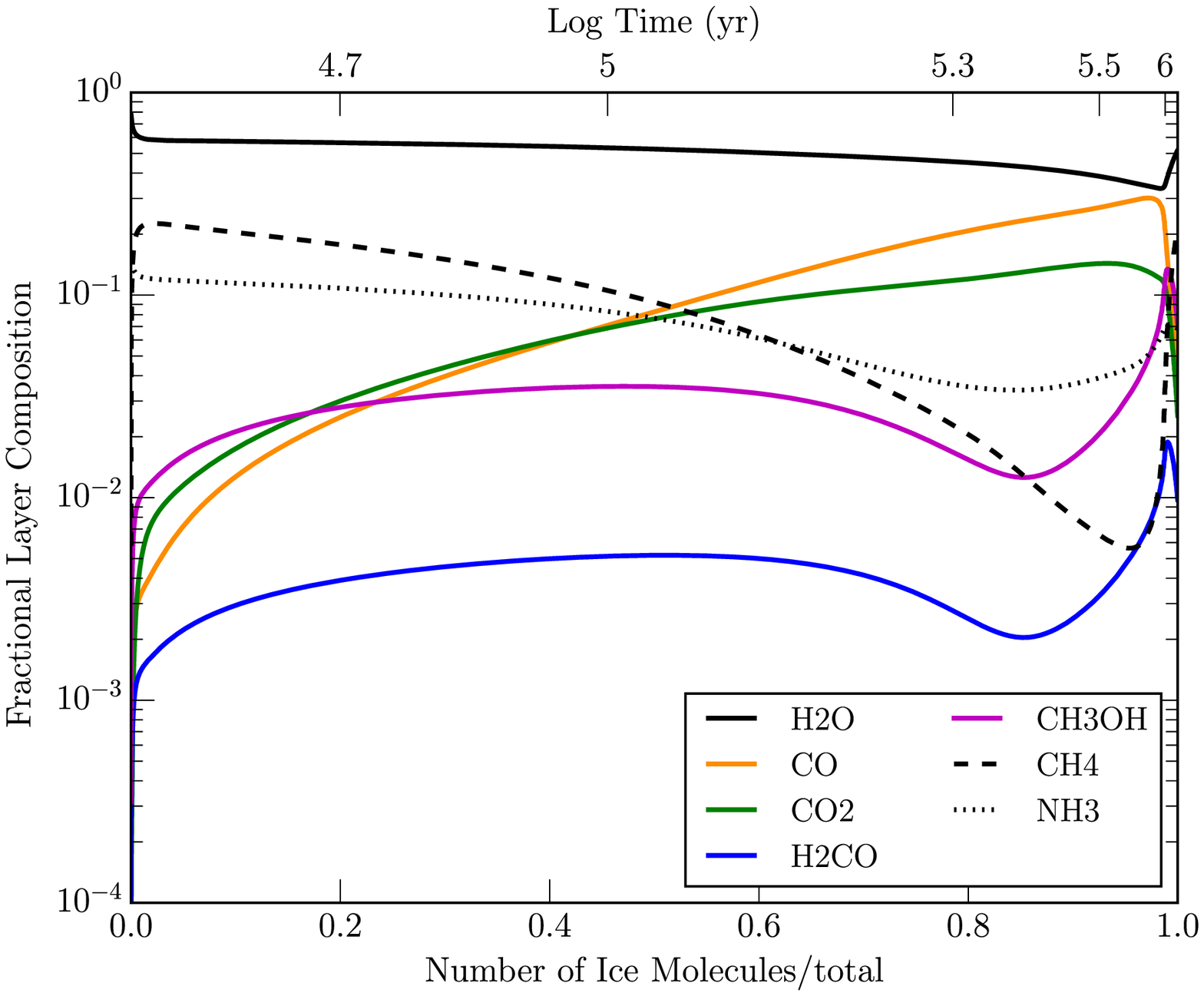}
        }
        \subfigure[]{
        }
     \end{center}
     \caption{%
        Shown is the evolution of the grain mantle species for the single grain models. Fractional composition of each ice monolayer is shown on the y axis for a given monolayer depth shown on the x axis, as normalized to the final amount of ice in the mantle.   }%
    \label{fig:1Gchem}
\end{figure*}

Figure~\ref{fig:1Gchem} shows the depth-dependence of important mantle species for the three single-grain models. Ice abundances are plotted as a fraction of the total ice present in each monolayer, versus the depth of that monolayer, which is normalized to the final amount of ice in the mantle at the end of the model run. Plotted values toward the left of the plot indicate ices that were deposited earlier in the chemical evolution of the cloud, and which are thus deeper down in the mantles; values toward the right were more recently deposited and nearer to the surface. $\mathrm{H_2O}$ comprises the majority of the ice. CO is gradually formed in the gas phase and accreted onto the surface; its abundance in the ice grows steadily with its gas-phase abundance.  CO$_2$ is produced on the surface; the dominant formation mechanism in this model is formation of OH on a CO ice surface, as described by \citet{Garrod11}.  This produces a final CO$_2$ abundance of roughly 25 percent with respect to CO for the 8 and 10 K cases, although it is somewhat higher in the 12 K model. These ratios are low when compared with observed values for dark clouds.

Shown in Table~\ref{table:mantlefraction} are the mantle ice fractions with respect to water at $10^6$ years for all models. This time is chosen for the model output as it is a representative time scale for the quiescent stage of molecular clouds. The CO ice abundance matches well with observed dark cloud values \citep[e.g. Elias 16, from][]{Gibb00}. High-mass young stellar object observations from \citet{Gibb00} and \citet{Whittet11} show a range of abundances with varying levels of model agreement. CO$_2$ abundance in the three single grain models increases with increasing dust temperature. This reproduces the results of prior work - near 12 K, CO surface diffusion and reaction with OH becomes competitive with the water-forming reaction OH + $\mathrm{H_2}$, efficiently converting CO to $\mathrm{CO_2}$. 

The hydrogen-rich species $\mathrm{CH_4}$ and $\mathrm{NH_3}$ are strongly produced in the model at early times, with abundances significantly greater than observed values.  This could be a result of atomic initial conditions and the lack of treatment for the cloud's early formation period.  $\mathrm{CH_4}$ is more abundant in low temperature models, likely a result of decreased reaction competition for carbon hydrogenation at lower temperatures (see also \citet{Garrod11} on this point).

\begin{deluxetable}{lrrrrrrr}
\tablecolumns{8}
\tablewidth{0pc}
\tablecaption{Mantle abundances at $\mathrm{5 \times 10^6}$ years for relevant species in models and observed dark clouds. The value of $\mathrm{H_2O}$ listed is with respect to total atomic hydrogen abundance, while the values for other species are given as abundance in percent of the $\mathrm{H_2O}$ mantle abundance. For the two observations, the $\mathrm{H_2O}$ value is the observed column density, with the following column values in similar fashion to the models.$^a$: \citet{Gibb00}; $^b$: \citet{Whittet11}}
\tablehead{
\colhead{Species} & \colhead{$\mathrm{H_2O}$}   & \colhead{$\mathrm{CO}$}    & \colhead{$\mathrm{CO_2}$}   &
\colhead{$\mathrm{CH_3OH}$}  & \colhead{$\mathrm{H_2CO}$} & \colhead{$\mathrm{CH_4}$}  & \colhead{$\mathrm{NH_3}$}}
\startdata
 1G\_T8     & 	2.08(-4) & 21.6 & 5.0 & 6.2 & 1.0 & 24.4 & 14.1 \\
 1G\_T10   & 	2.02(-4) & 23.1 & 6.7 &	5.8 & 0.89 & 23.6 & 14.3 \\
 1G\_T12   & 	1.84(-4) & 21.6 & 14.9 & 5.6 & 0.83 & 19.6 & 14.8 \\
 \hline
 5G\_T8\_UNIF  &	2.16(-4) & 19.1 & 4.6 & 5.5 & 0.92 & 27.1 & 14.9 \\
 5G\_T10\_UNIF & 	2.03(-4) & 23.2 & 6.8 & 5.5 & 0.85 & 23.4 & 15.1 \\
 5G\_T12\_UNIF  & 	1.83(-4) & 21.9 & 15.3 & 5.4 & 0.80 & 19.9 & 15.2 \\
 \hline
 5G\_T8\_DIST   &	2.02(-4) & 23.7 & 7.1 & 5.7 & 0.88 & 22.5 & 15.3 \\
 5G\_T10\_DIST   & 	1.59(-4) & 7.9 & 36.2 & 2.9 & 0.42 & 23.6 & 20.5 \\
 5G\_T12\_DIST   & 	1.58(-4) & 1.1 & 43.2 & 2.1 & 0.32 & 25.6 & 26.8 \\
 \hline
 \vspace{0.3cm}
 5G\_Coll   & 	1.39(-4) & 29.7 & 31.0 & 6.3 & 1.1 & 12.4 & 21.7 \\
 Elias 16$^a$   &  	2.5(18) & 25 & 18 & \textless 3 & ... & ... & $\leq$ 9 \\
 W33A$^a$     &	1.10(19) & 8 & 13 & 18 & 3 & 4 & 15 \\
 NGC 7538 IRS9$^b$ & 6.7(18) & 37.6 & 23.1 & 7.1 & ... & ... & ... \\
\enddata
\label{table:mantlefraction}
\end{deluxetable}

Methanol ($\mathrm{CH_3OH}$) is produced by the successive addition of hydrogen to CO on the grain surface. Efficient production of methanol requires an availability of atomic hydrogen and sufficient time for it to react with CO and $\mathrm{H_2CO}$ - the limiting steps in the formation pathway, due to the activation energy barriers present in those reactions.  If accretion, and consequent formation of new ice-mantle layers, is too rapid, CO and $\mathrm{H_2CO}$ can be frozen into the mantle without full conversion to methanol. Methanol sees fairly consistent ice formation throughout the single grain models.

\subsection[]{Chemical Evolution - Grain Size Distribution}

\begin{figure*}
     \begin{center}
        \subfigure[Aggregate Ice Mantle]{%
            \label{fig:5g10unifgr0}
            \includegraphics[width=0.46\textwidth]{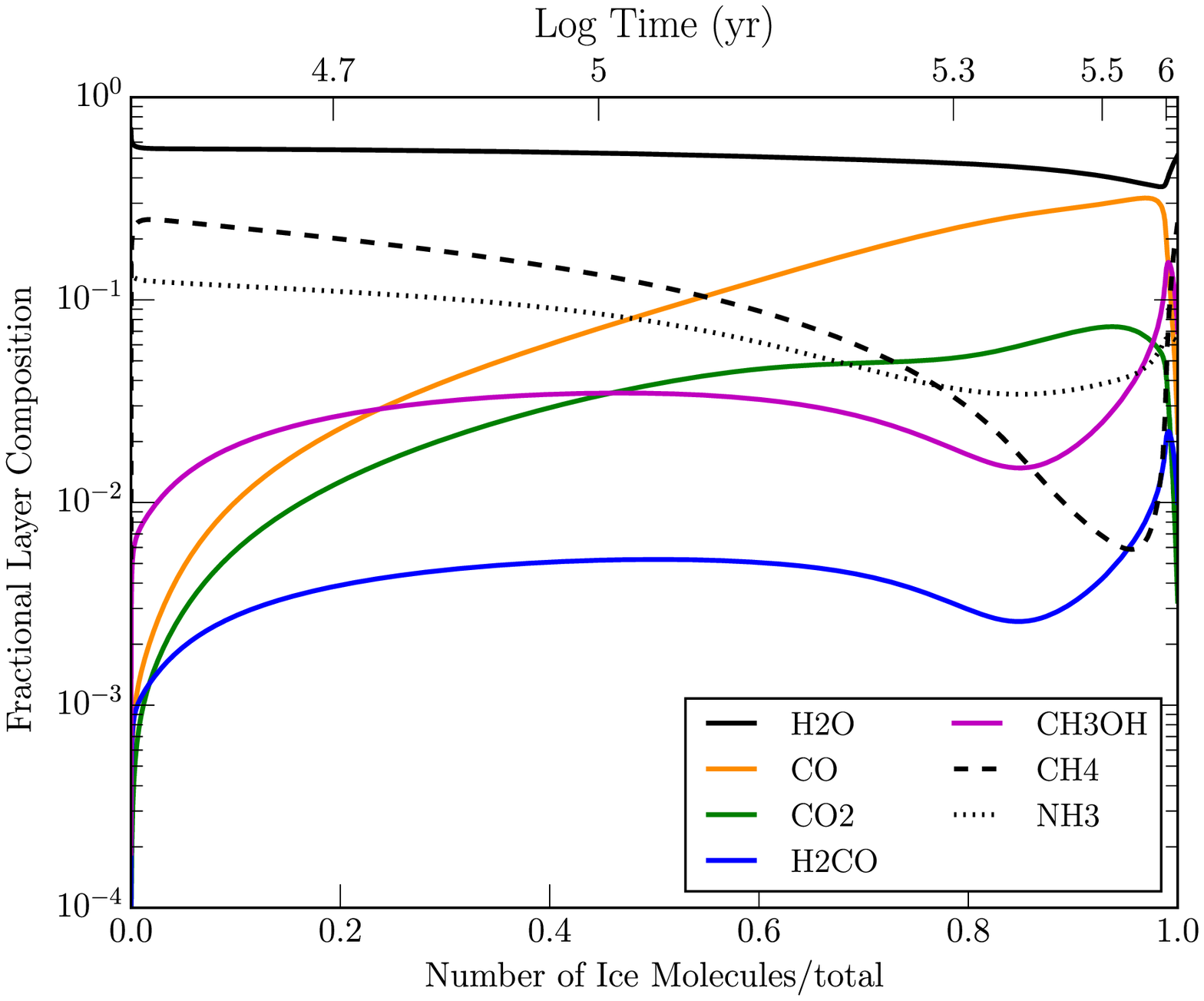}
        }%
        \subfigure[Grain Size 1]{%
           \label{fig:5g10unifgr1}
           \includegraphics[width=0.46\textwidth]{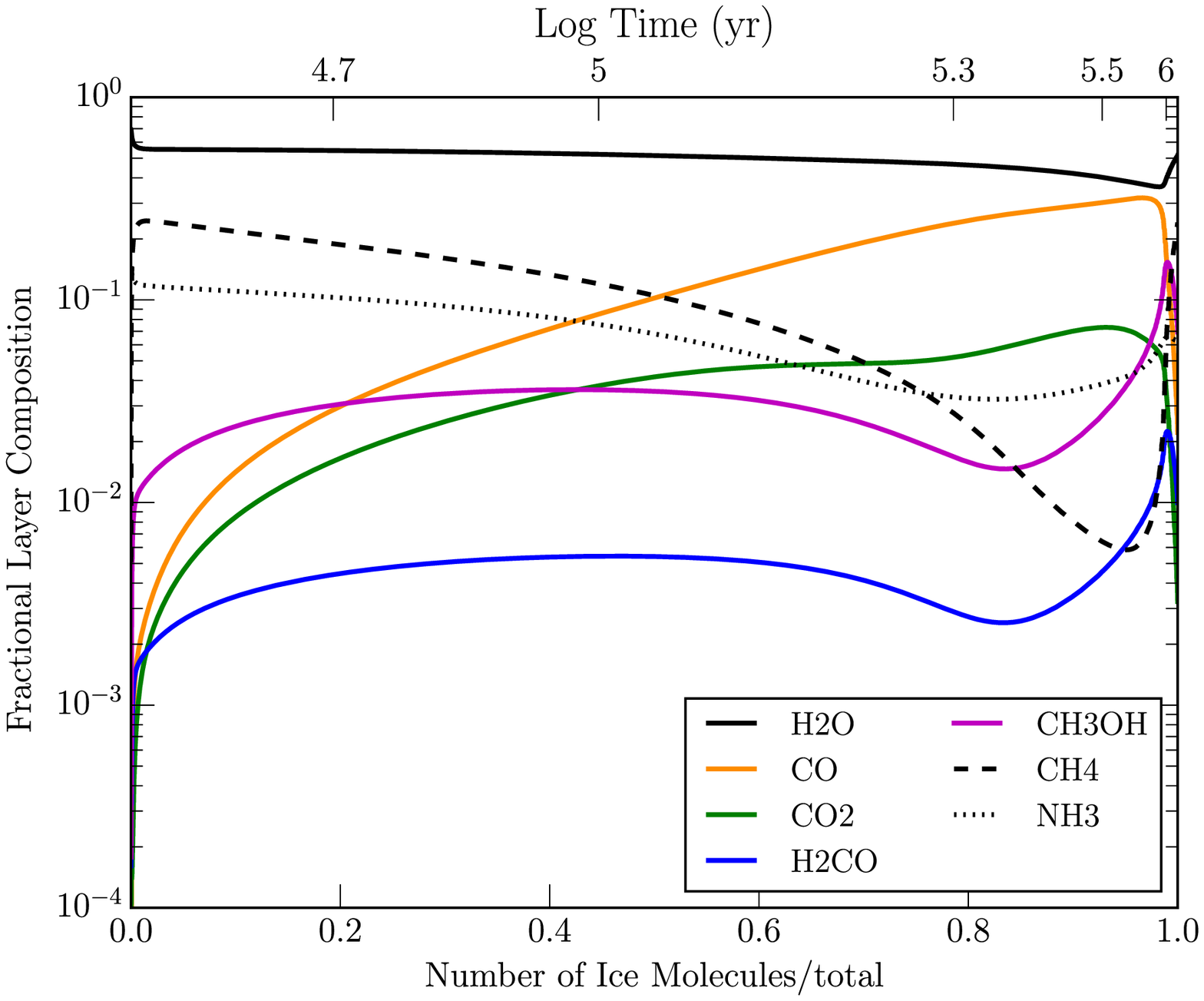}
        }\\ 
        \subfigure[Grain Size 2]{%
           \label{fig:5g10unifgr2}
           \includegraphics[width=0.46\textwidth]{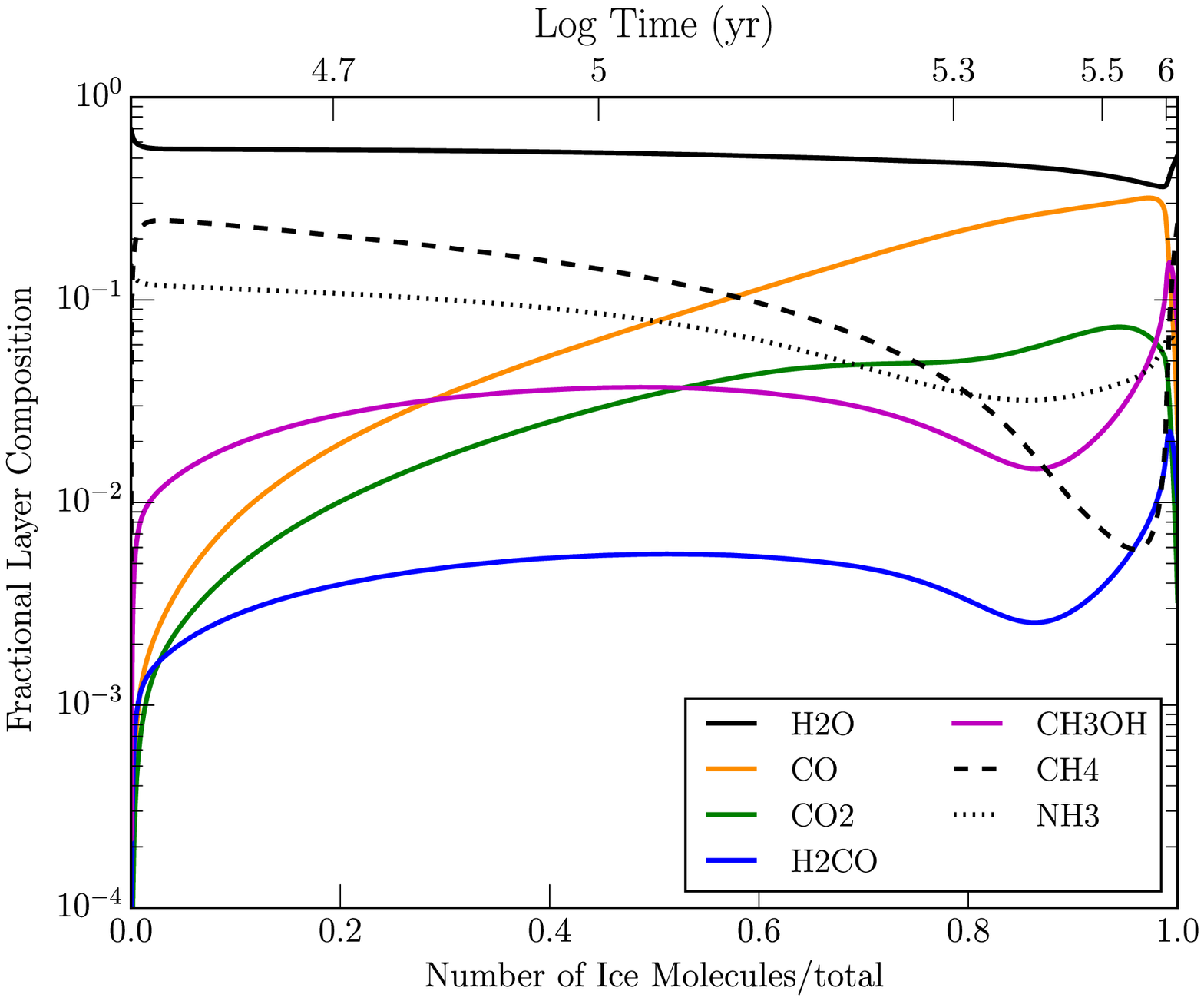}
        }%
        \subfigure[Grain Size 3]{%
           \label{fig:5g10unifgr3}
           \includegraphics[width=0.46\textwidth]{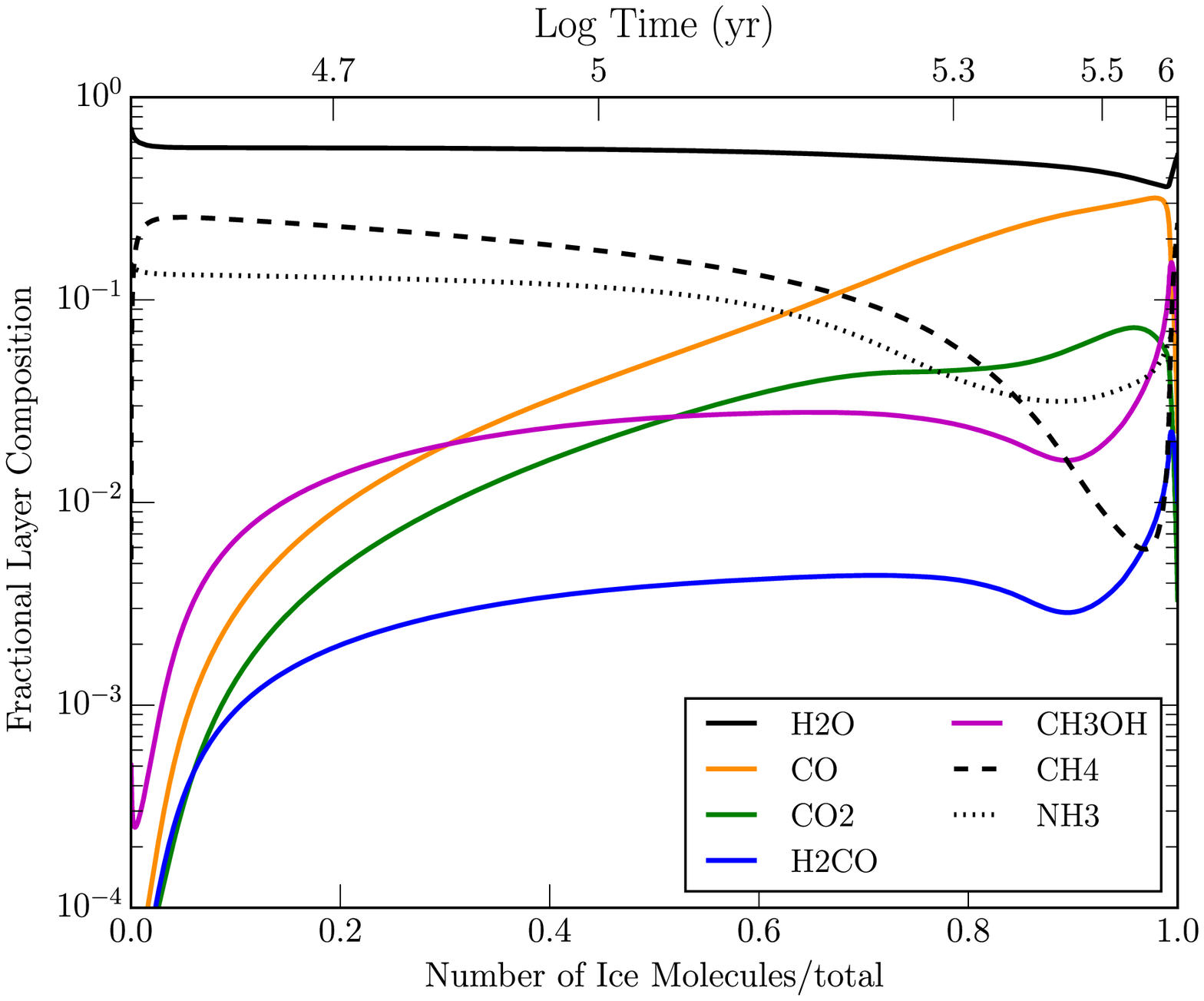}
        }\\ 
        \subfigure[Grain Size 4]{%
            \label{fig:5g10unifgr4}
            \includegraphics[width=0.46\textwidth]{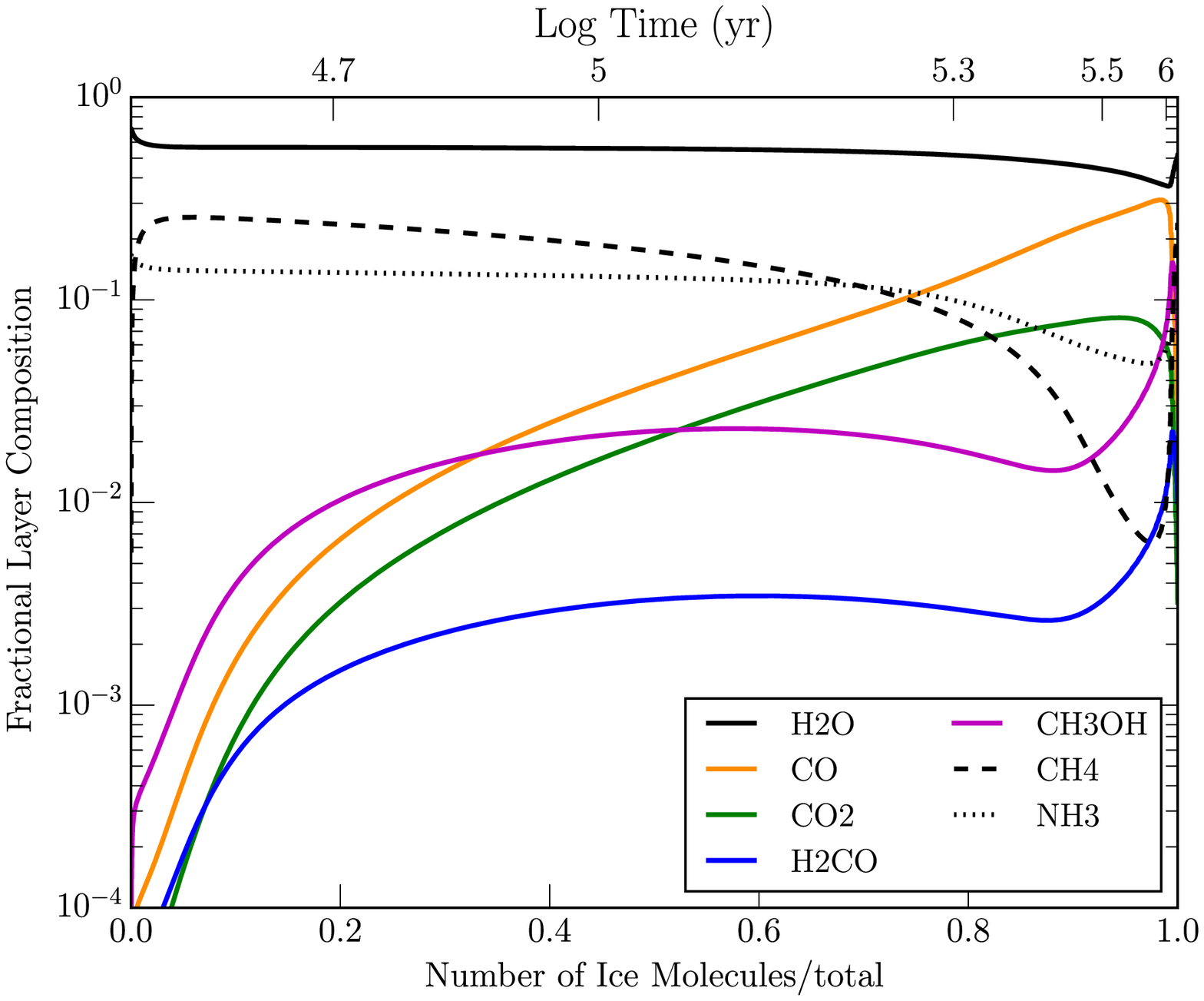}
        }%
        \subfigure[Grain Size 5]{%
            \label{fig:5g10unifgr5}
            \includegraphics[width=0.46\textwidth]{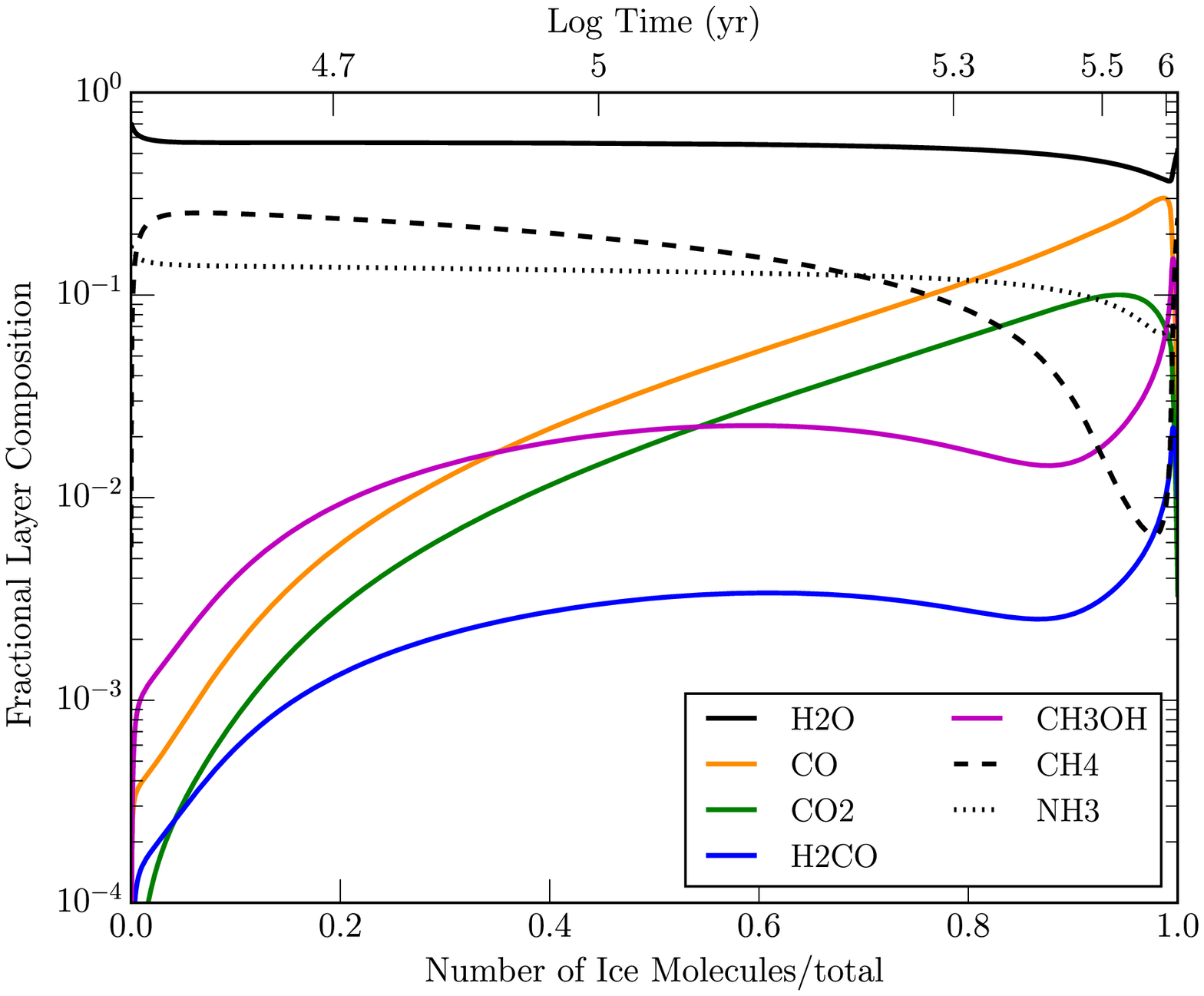}
        }%
    \end{center}
    \caption{%
        Chemical evolution for model 5G\_T10\_UNIF. The upper left panel shows the evolution of the aggregate ice mantle, while the other panels show the evolution for the smallest grain (1) to the largest grain (5).  No large differences are present between the grain sizes.
     }%
   \label{fig:5g10unif}
\end{figure*}

The first implementation of the grain-size distribution begins with models of multiple grain sizes all fixed at a uniform temperature; separate models are run at 8, 10, and 12 K. All models begin with the same total grain cross-section. The temperature is held fixed throughout the model run, despite the differing initial grain size and the changing grain size through time, as seen in Figure~\ref{fig:graingrowth}. Differences in final composition between single-grain and multiple-grain models with equal temperature (e.g. 1G\_T8 and 5G\_T8\_UNIF) should be a result only of how the cross-sectional and surface area are distributed amongst the grains. Shown in Figure~\ref{fig:5g10unif} is the mantle evolution for model 5G\_T10\_UNIF.

\begin{figure*}
     \begin{center}
        \subfigure[Aggregate Ice - Model 5G\_T8\_UNIF]{
            \label{fig:first}
            \includegraphics[width=0.46\textwidth]{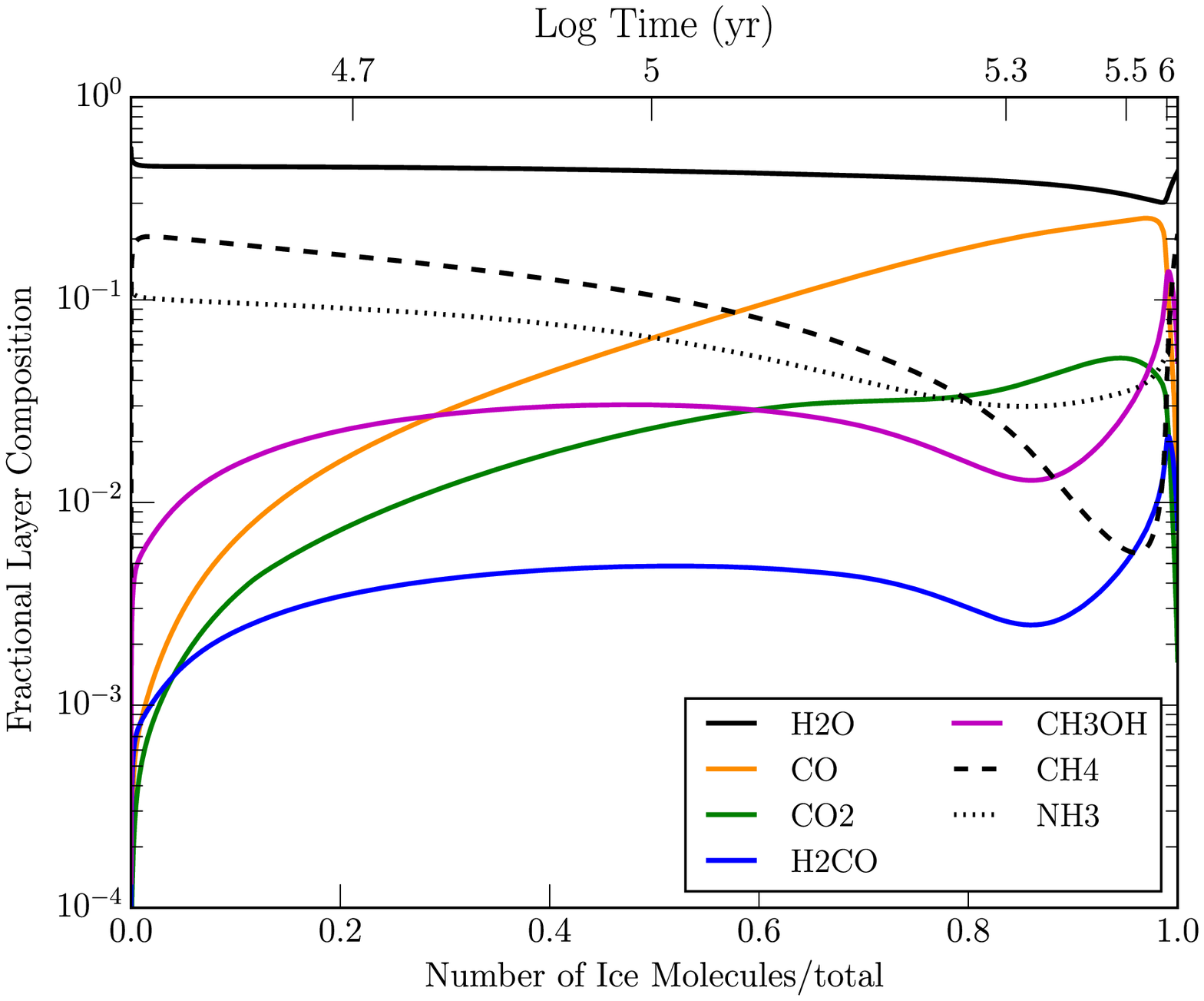}
        }
        \subfigure[Aggregate Ice - Model 5G\_T12\_UNIF]{
           \label{fig:second}
           \includegraphics[width=0.46\textwidth]{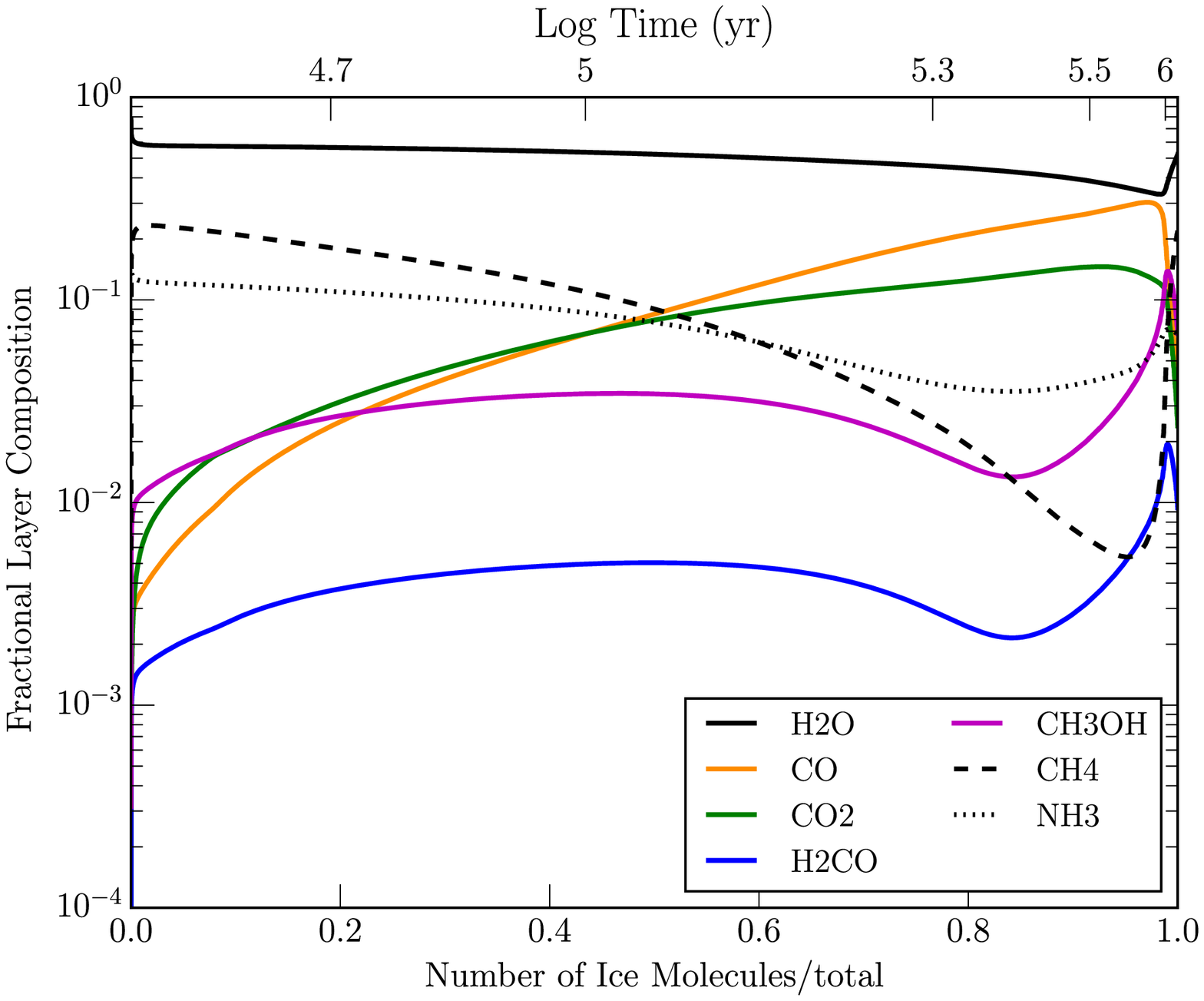}
        }
    \end{center}
    \caption{
        Aggregate mantle evolution for models 5G\_T8\_UNIF and 5G\_T12\_UNIF. 
     }
   \label{fig:extraunifaggregate}
\end{figure*}

For models with multiple grains, it can be informative to plot the aggregate behavior of all grain surface and mantle species. We plot the aggregate mantle behavior by weighting each individual grain surface abundance by that grain's accretion rate. We then sum the weighted abundances from each grain to find an aggregate surface value for a given species. The composition of new mantle ice follows from the composition of the surface. The aggregate mantle in Panel~\ref{fig:5g10unifgr0} shows little difference from the single grain model.  Comparing 1G\_T10 and 5G\_T10\_UNIF in Table~\ref{table:mantlefraction} shows the models are largely similar, with only a marginal discrepancy in $\mathrm{NH_3}$ abundance.  Examining the individual grain populations in 5G\_T10\_UNIF, the largest grains seem to produce more $\mathrm{NH_3}$ at later times than expected. Apart from $\mathrm{NH_3}$, it seems that the grain distribution alone has no large effect on the chemistry of the cloud.

Figure~\ref{fig:extraunifaggregate} shows the aggregate ice mantles for the 5G\_T8\_UNIF and 5G\_T12\_UNIF models, which constitute the sum of each species over all grain-size populations (naturally weighted according to the abundance of each grain population). CO and CO$_2$ abundances vary with temperature as expected - there is more CO$_2$ production and less CO at higher temperatures. 

The grain-size distribution introduces an individual accretion rate for each grain size, as well as a size-dependent total number of binding sites.  The change in number of binding sites can affect the chemistry by changing the amount of time taken for a species to diffuse across the full grain surface. However, the aggregate behavior of the five-grain, uniform-temperature models is comparable to the single-grain models, implying that a grain-size distribution alone does not strongly affect the surface chemistry. 

\begin{figure*}
     \begin{center}
        \subfigure[Aggregate Ice Mantle]{%
            \label{fig:5gt10distgr0}
            \includegraphics[width=0.46\textwidth]{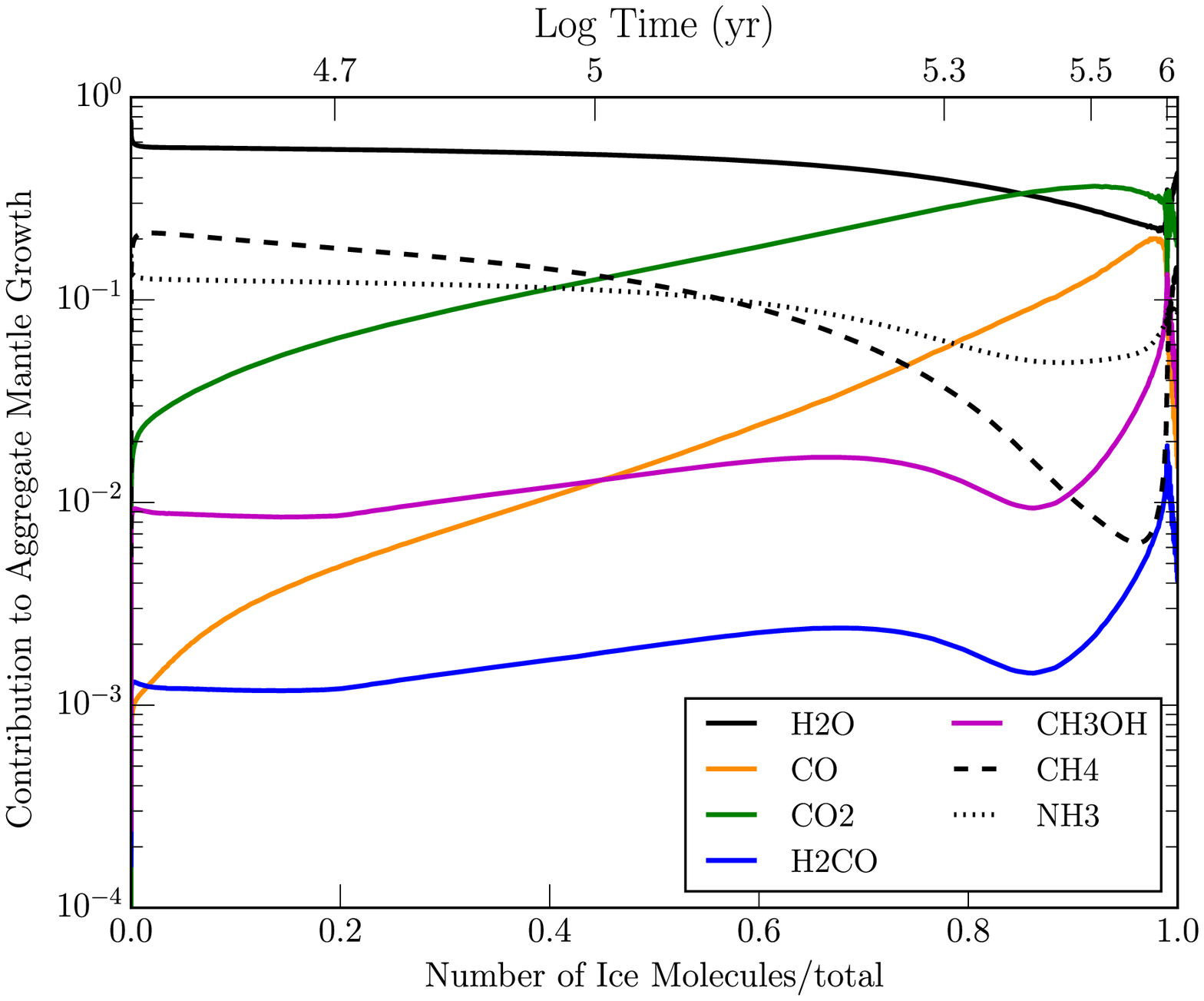}
        }%
        \subfigure[Grain Size 1]{%
           \label{fig:5gt10distgr1}
           \includegraphics[width=0.46\textwidth]{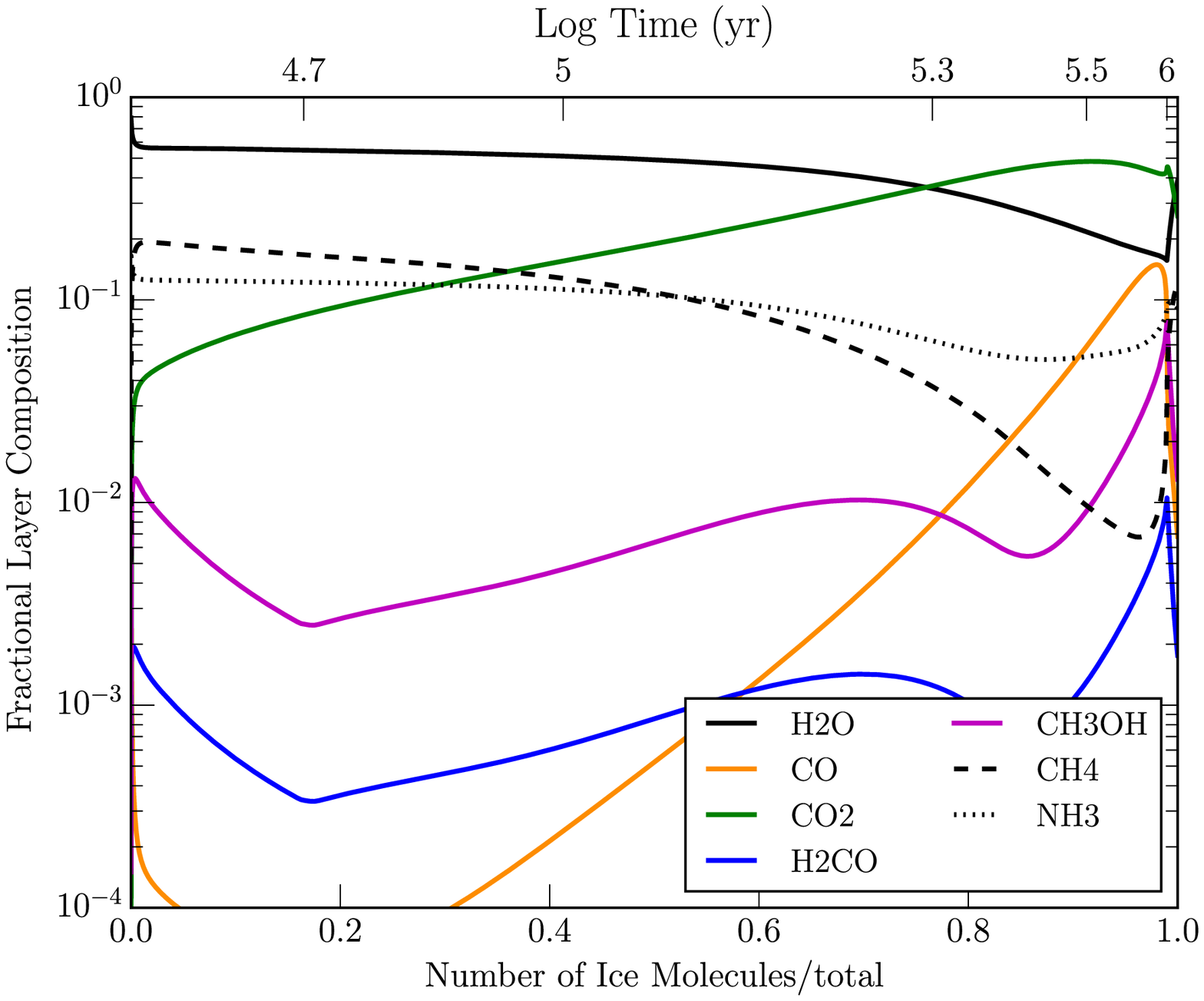}
        }\\ 
        \subfigure[Grain Size 2]{%
            \label{fig:5gt10distgr2}
            \includegraphics[width=0.46\textwidth]{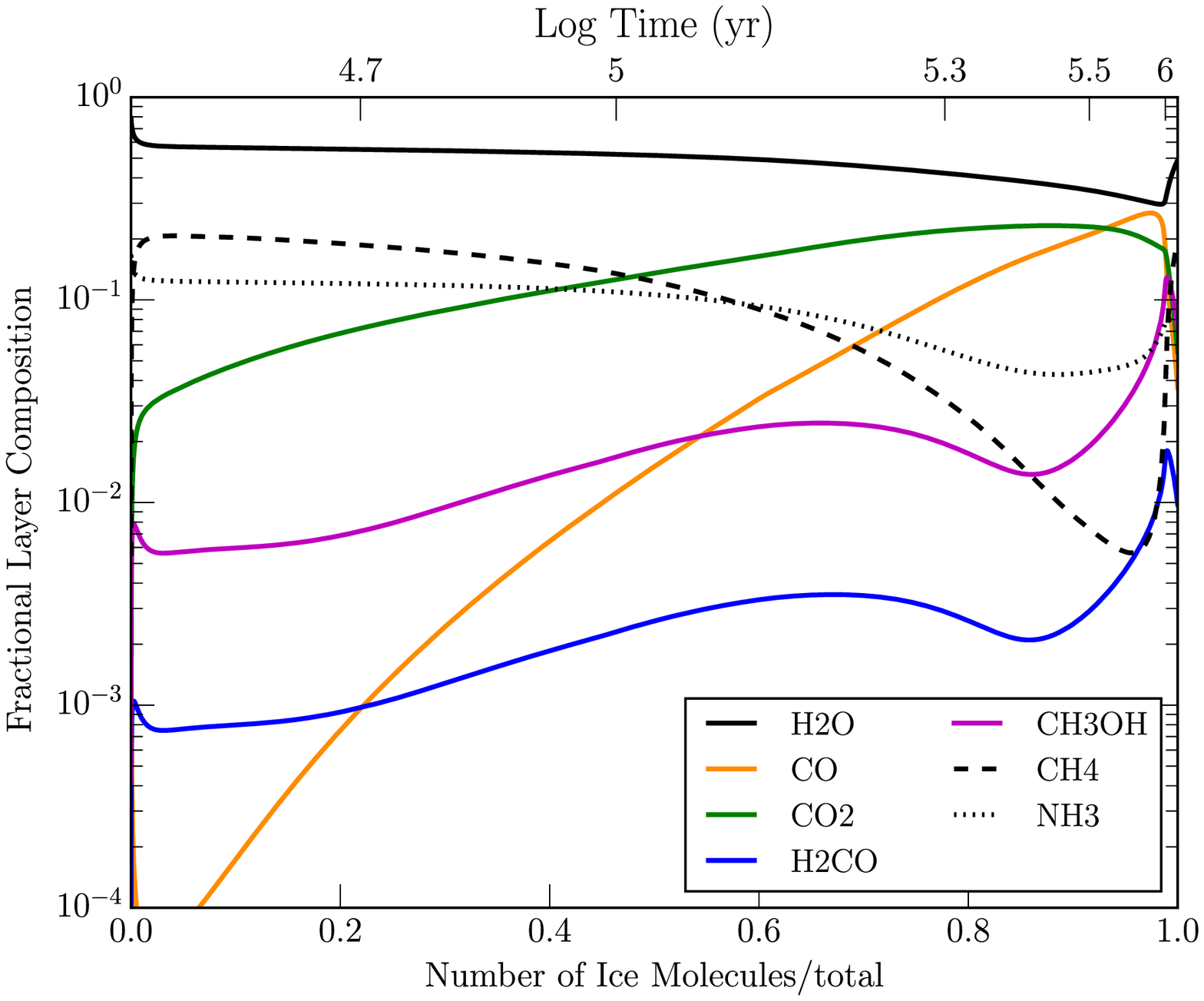}
        }%
        \subfigure[Grain Size 3]{%
           \label{fig:5gt10distgr3}
           \includegraphics[width=0.46\textwidth]{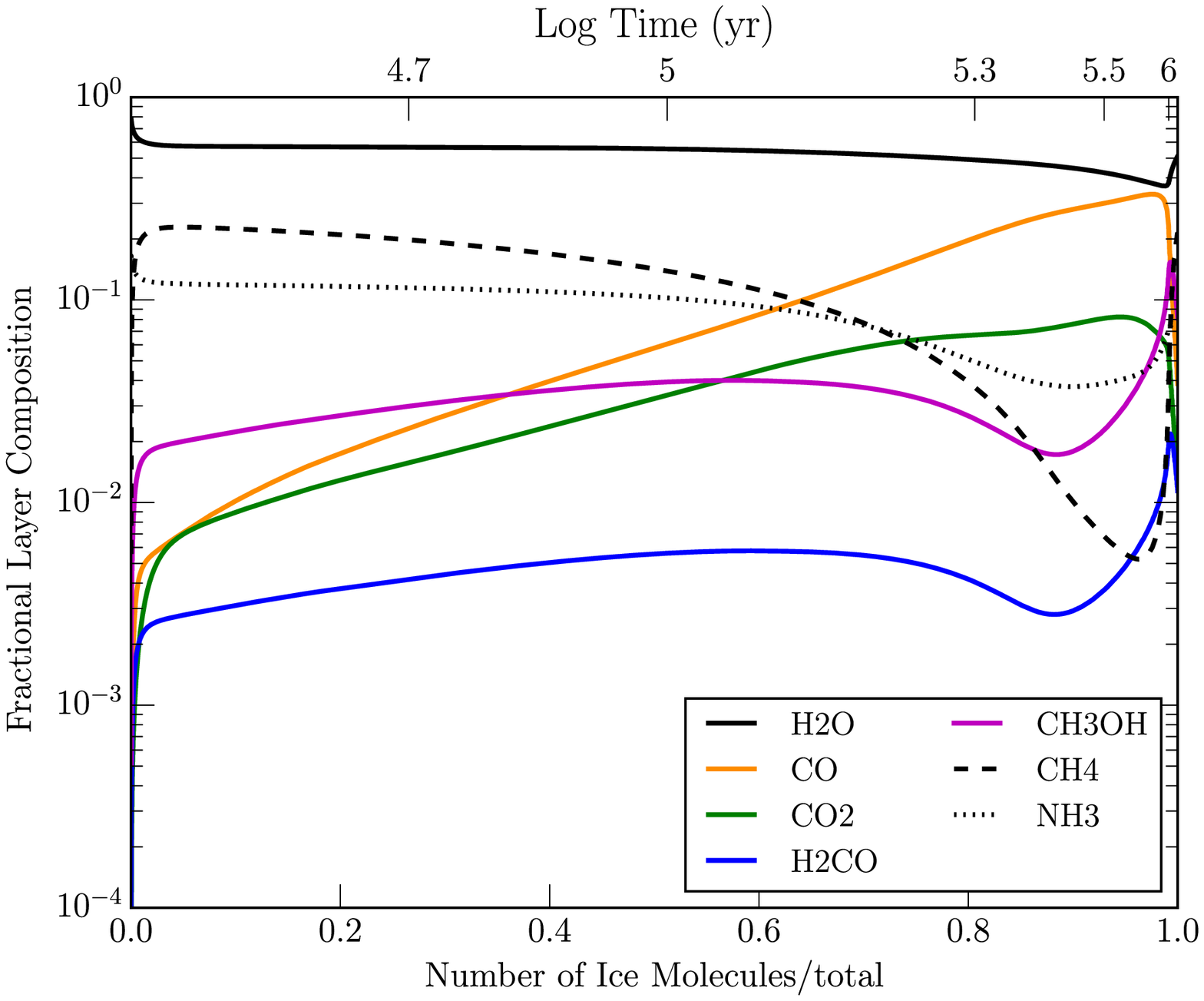}
        }\\ 
        \subfigure[Grain Size 4]{%
            \label{fig:5gt10distgr4}
            \includegraphics[width=0.46\textwidth]{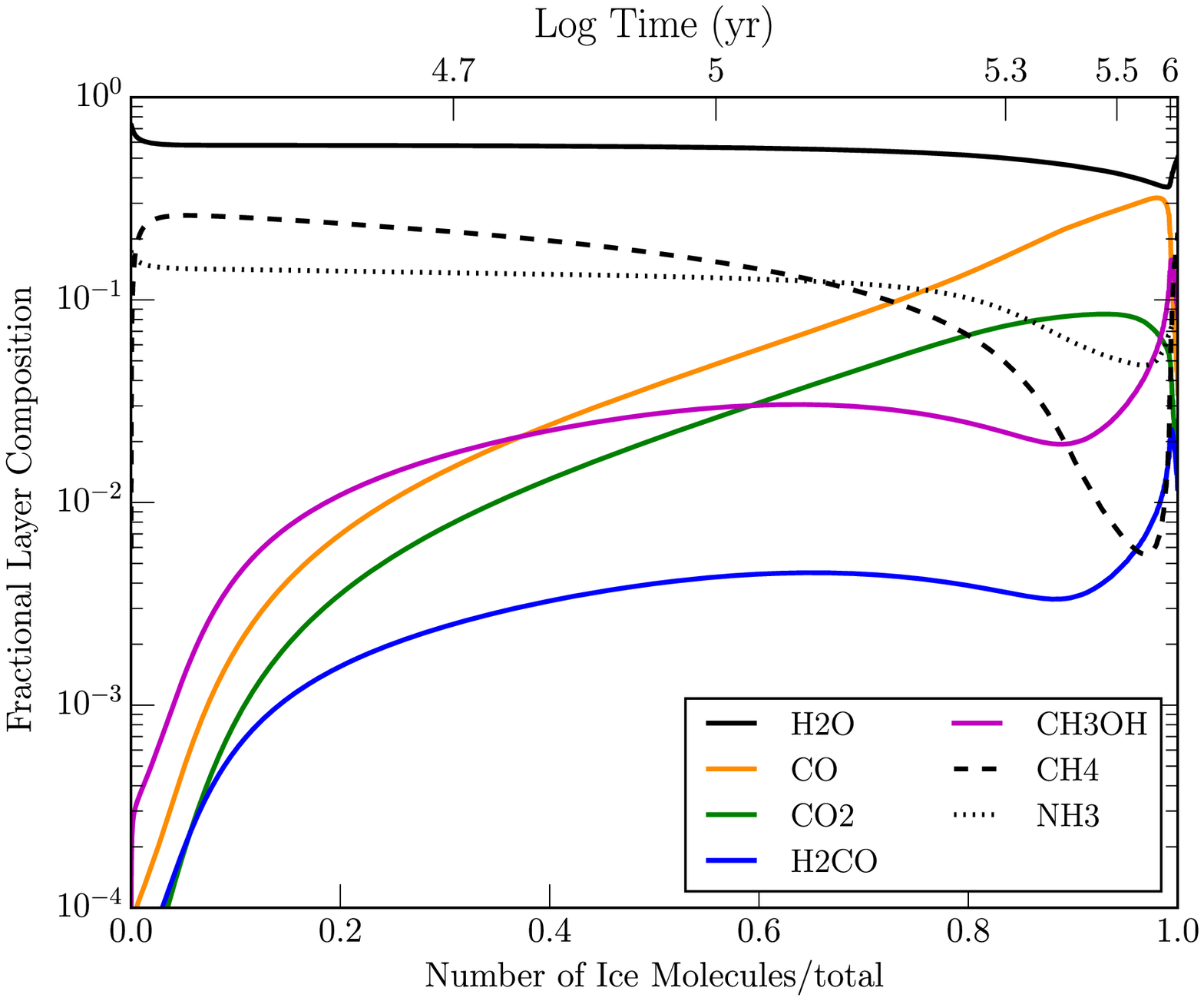}
        }%
        \subfigure[Grain Size 5]{%
            \label{fig:5gt10distgr5}
            \includegraphics[width=0.46\textwidth]{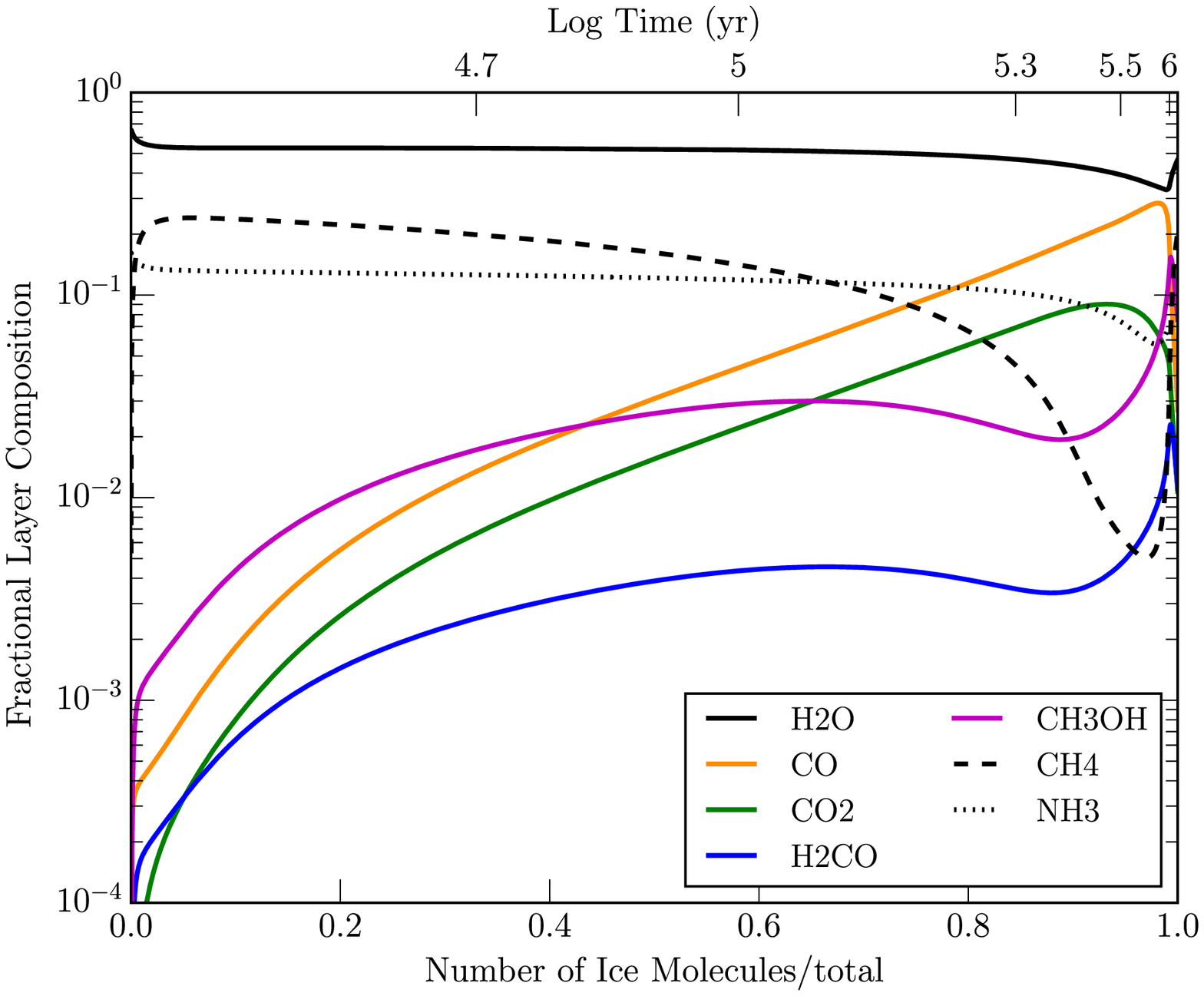}
        }%
    \end{center}
    \caption{%
        Chemical evolution for model 5G\_T10\_DIST. The upper left panel shows the evolution of the aggregate ice mantle, while the other panels show the evolution for the smallest grain (1) to the largest grain (5).  The temperature variation, both between grains and as individual grain populations grow, determine the CO/$\mathrm{CO_2}$ chemistry.
     }%
   \label{fig:5gt10dist}
\end{figure*}

\subsection[]{Chemical Evolution - Grain Size and Temperature Distribution}

Figure~\ref{fig:5gt10dist} shows the evolution of the mantle chemistry for model 5G\_T10\_DIST, in which the temperature of each grain population varies with the combined grain and ice mantle radius. It is apparent that the initial dust temperature distribution greatly affects the surface chemistry. On the two smallest grains, shown in Panels~\ref{fig:5gt10distgr1} and \ref{fig:5gt10distgr2}, CO$_2$ formation is initially highly efficient, with initial $\mathrm{T_d} \ge 12 K$. On the larger grains, the cold chemistry remains, and CO is more abundant than $\mathrm{CO_2}$.  As shown in Table~\ref{table:mantlefraction}, the total aggregate amount of water ice has decreased due to increased competition by CO in reactions involving OH. 

The effects of grain growth are also seen in the chemical evolution of the small grains. As the ice mantles grow and $\mathrm{T_d}$ drops (for grain size 1, from 15.6 K to 12.9 K), the $\mathrm{CO_2}$/CO ratio decreases.  The effect is most dramatic in the two smallest sizes, as CO fractional abundance changes by up to three orders of magnitude. Grain size 2 (Panel~\ref{fig:5gt10distgr2}) crosses the temperature threshold for efficient CO$_2$ production, and briefly has a surface $\mathrm{CO_2}$/CO ratio \textless 1. As a result of the efficient CO$_2$ production, aggregate CO abundance drops to only 8 percent with respect to $\mathrm{H_2O}$. This is on the low end of observed dark cloud values (Table 1).

The hydrogenated species in Model 5G\_T10\_DIST are again overproduced. $\mathrm{NH_3}$ abundance has increased when compared with model 5G\_T10\_UNIF and is greater than observed dark cloud values. $\mathrm{CH_4}$ abundance is relatively consistent across all static models. $\mathrm{CH_3OH}$ abundance is consistent with observed values and is not affected greatly by the temperature distribution.

\begin{figure*}
     \begin{center}
        \subfigure[5G\_T8\_DIST]{%
            \label{fig:first}
            \includegraphics[width=0.46\textwidth]{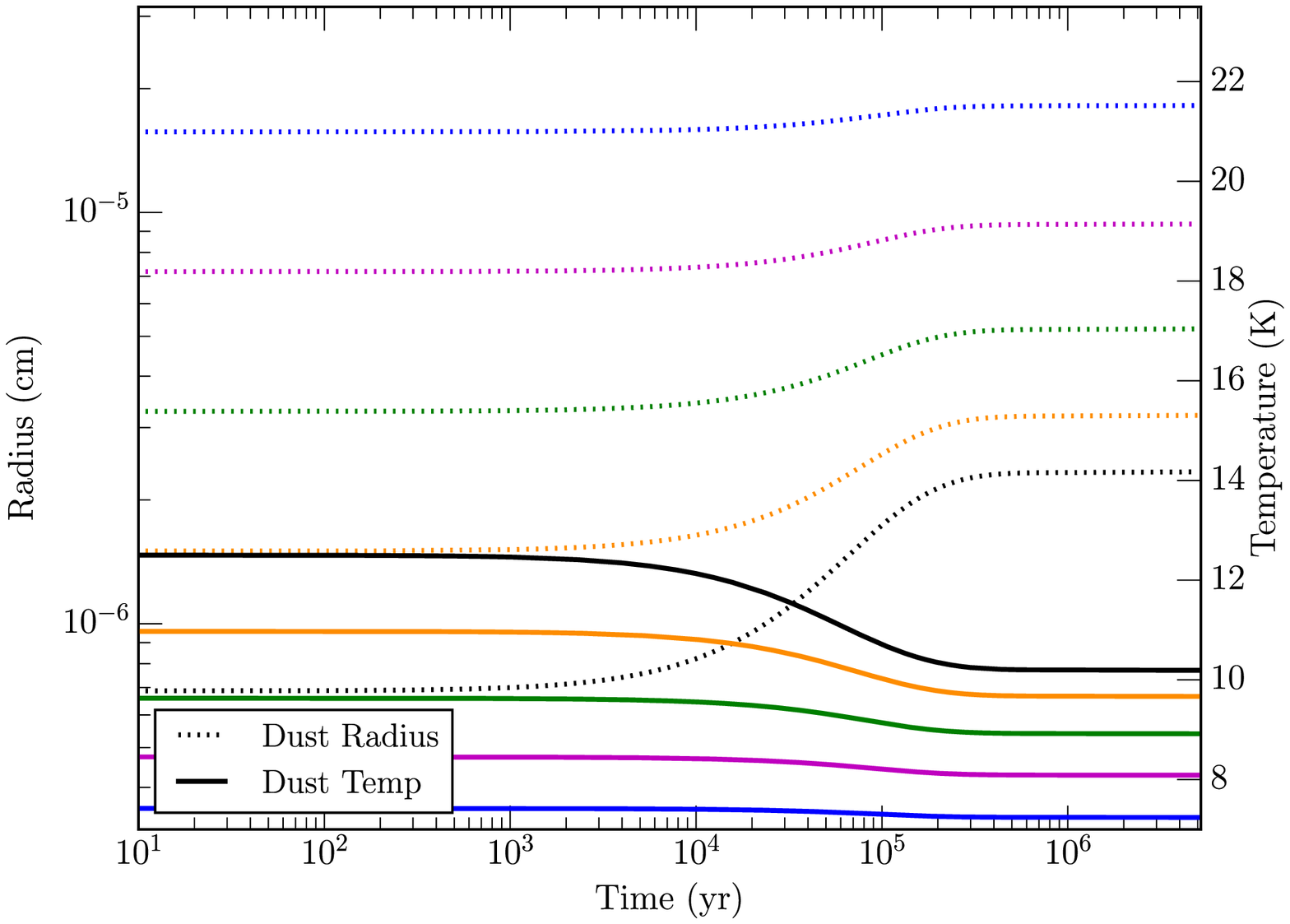}
        }%
        \subfigure[5G\_T12\_DIST]{%
           \label{fig:second}
           \includegraphics[width=0.46\textwidth]{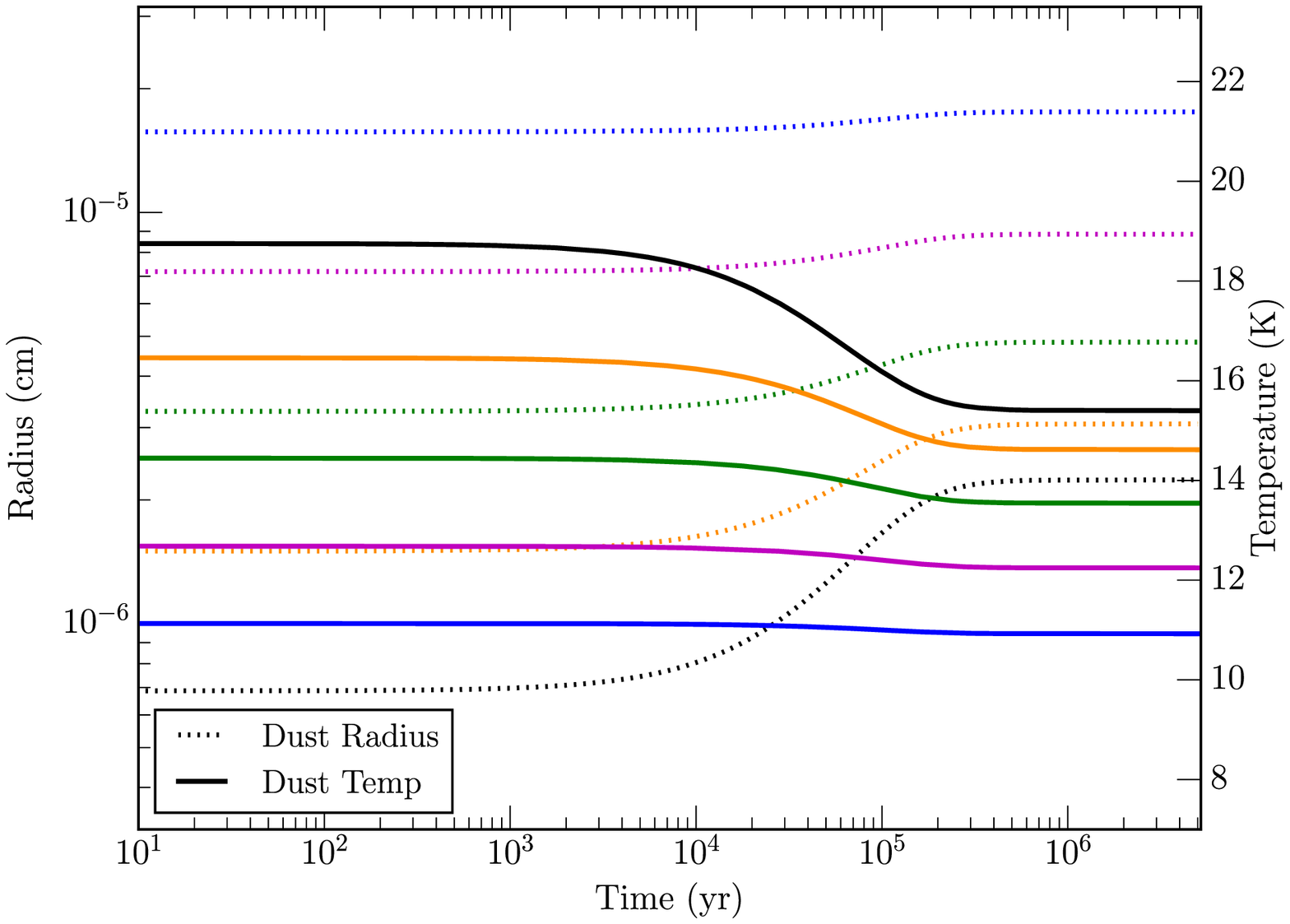}
        }
    \end{center}
    \caption{%
        Grain size and temperature evolution for the 8 K and 12 K models including a temperature distribution across five grain sizes. The size variation between the two models is fairly similar, indicating that the difference in temperature does not strongly affect accretion rates (and therefore temperature evolution, relative to the starting temperature.)
     }%
   \label{fig:extradistgrowth}
\end{figure*}

With a grain temperature distribution, the 8 K and 12 K models are more varied when compared with the 10 K equivalent model (again, with this temperature being assigned to the canonical grain with radius $0.1 \mathrm{\mu m}$). Figure~\ref{fig:extradistgrowth} shows the size and temperature evolution of models 5G\_T8\_DIST and 5G\_T12\_DIST. The mantle-growth behavior is comparable, while the general form of the temperature evolution is similar but scaled by the 4 K difference. This is expected, as the temperature evolution is determined by the rate of accretionary growth.

\begin{figure*}
     \begin{center}
        \subfigure[Aggregate Ice Mantle]{%
            \label{fig:first}
            \includegraphics[width=0.46\textwidth]{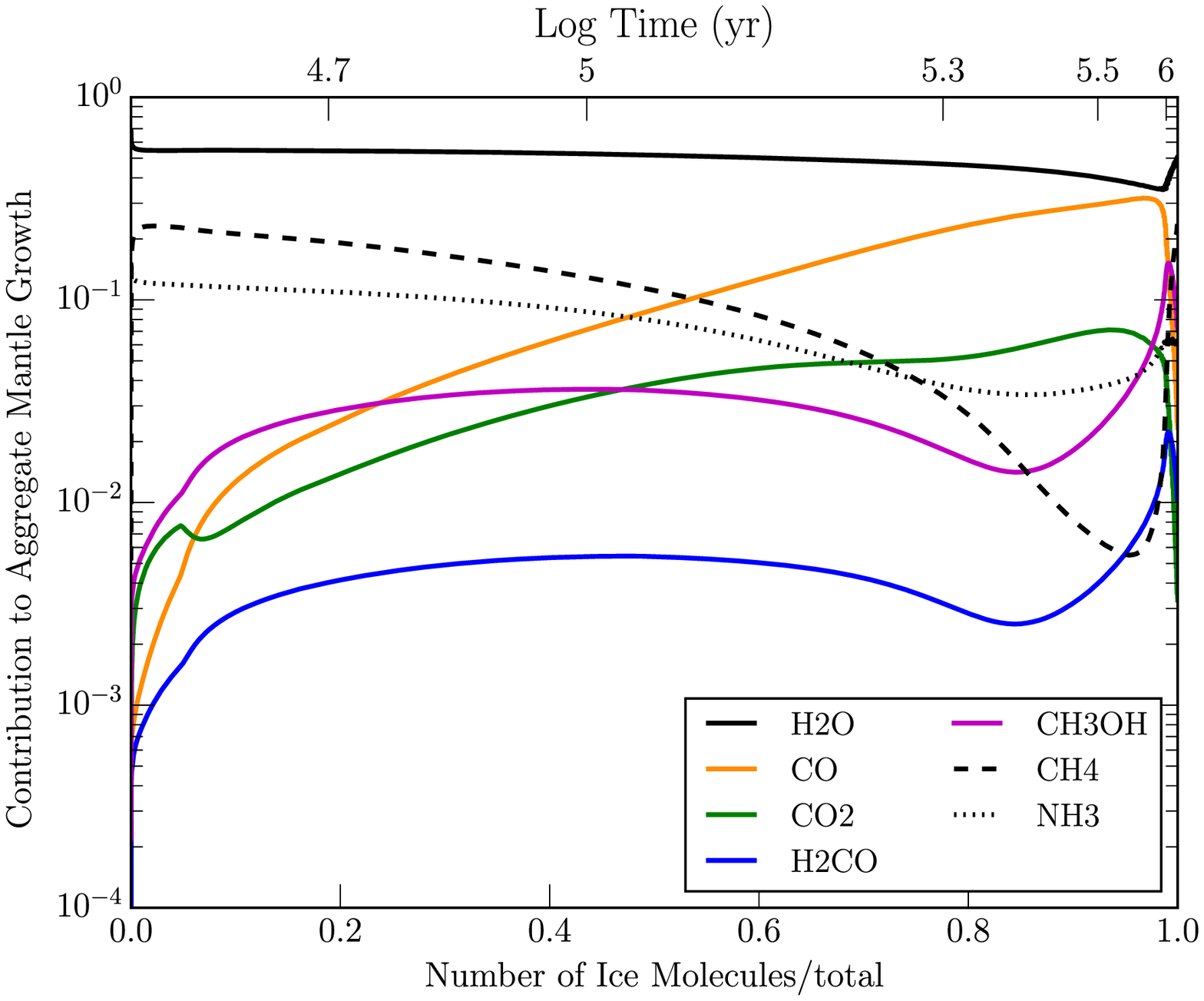}
        }%
        \subfigure[Aggregate Ice Mantle]{%
           \label{fig:second}
           \includegraphics[width=0.46\textwidth]{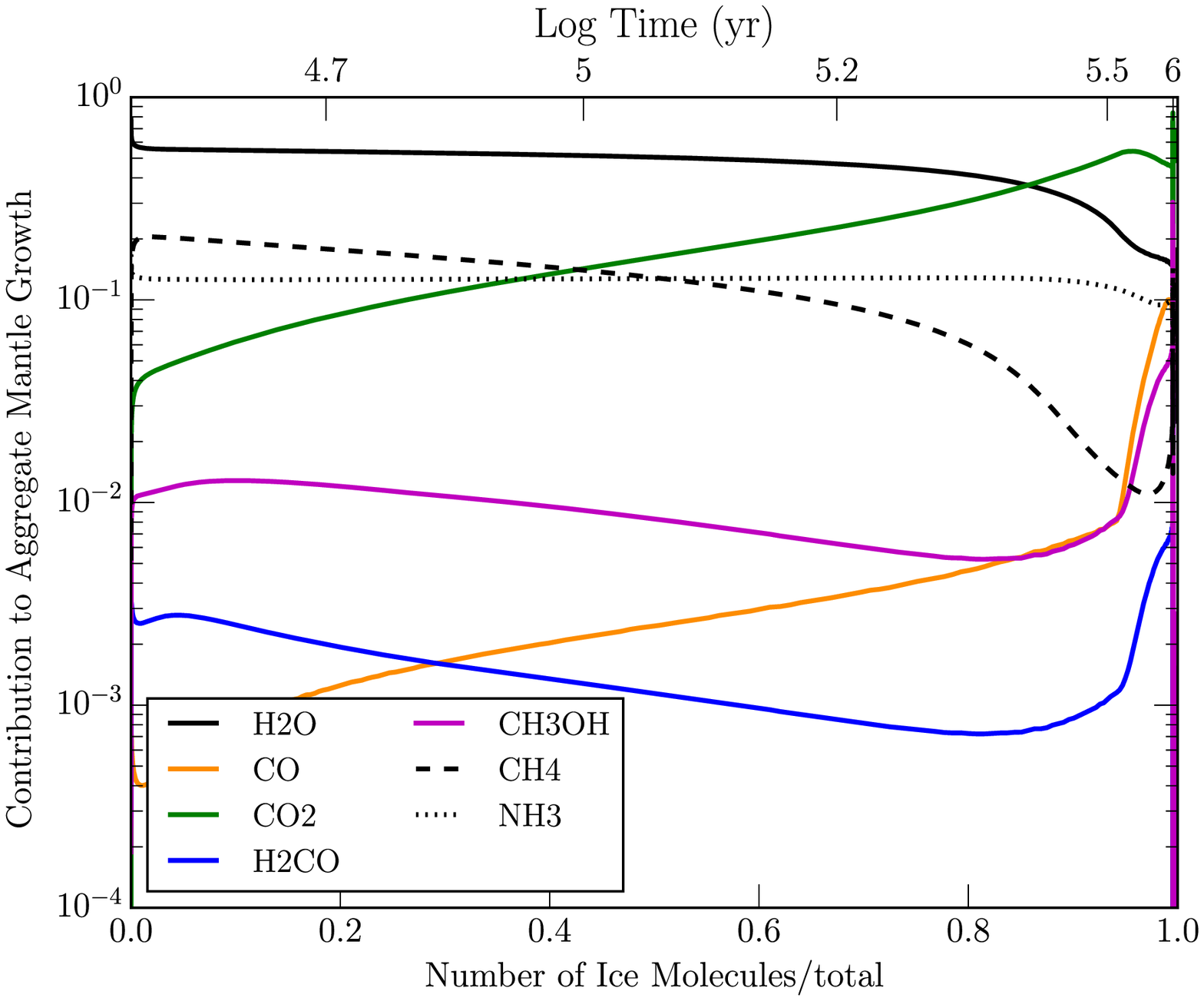}
        }\\ 
        \subfigure[Grain Size 1]{%
            \label{fig:third}
            \includegraphics[width=0.46\textwidth]{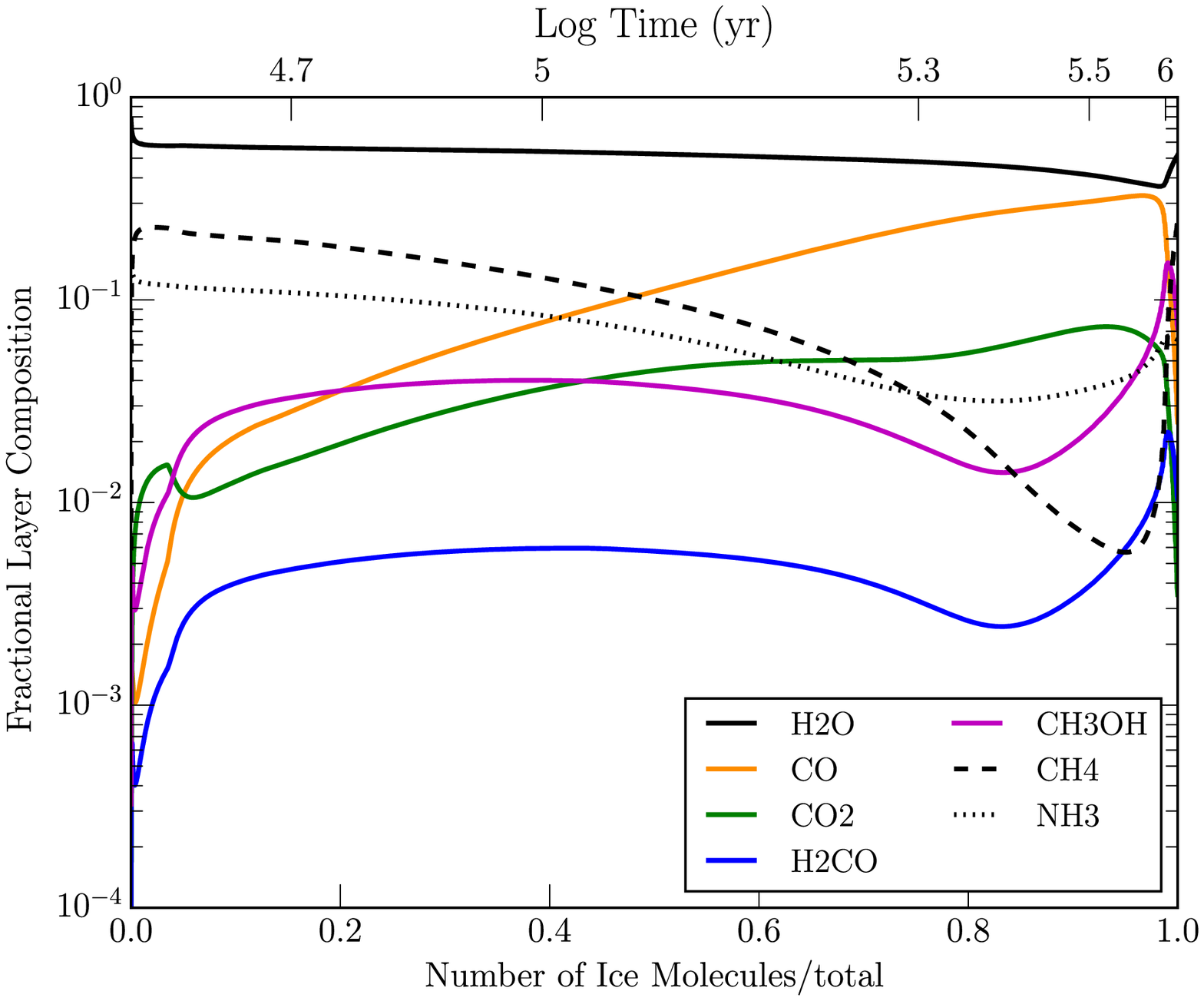}
        }%
        \subfigure[Grain Size 1]{%
            \label{fig:fourth}
            \includegraphics[width=0.46\textwidth]{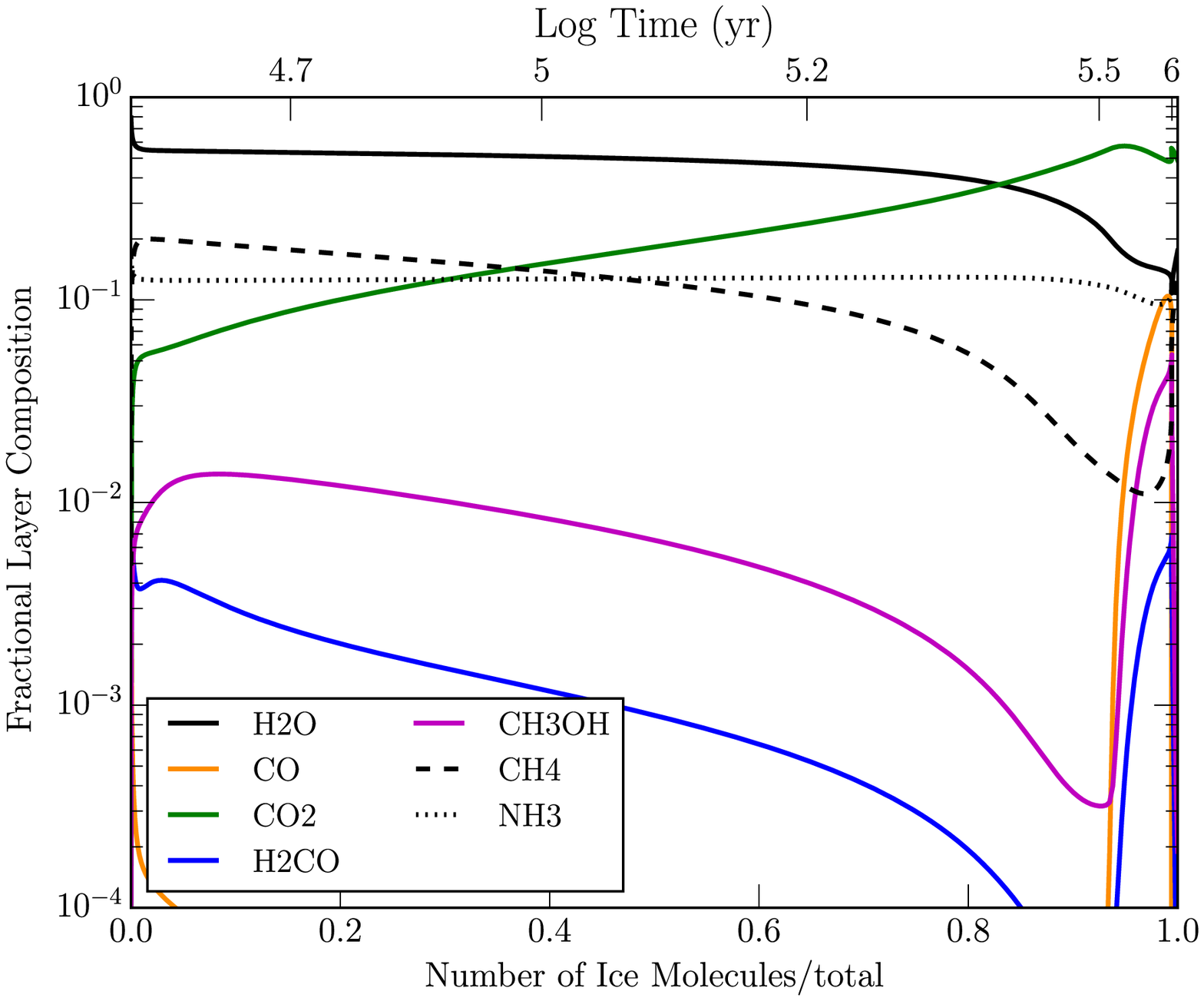}
        }\\ 
        \subfigure[Grain Size 5]{%
            \label{fig:fifth}
            \includegraphics[width=0.46\textwidth]{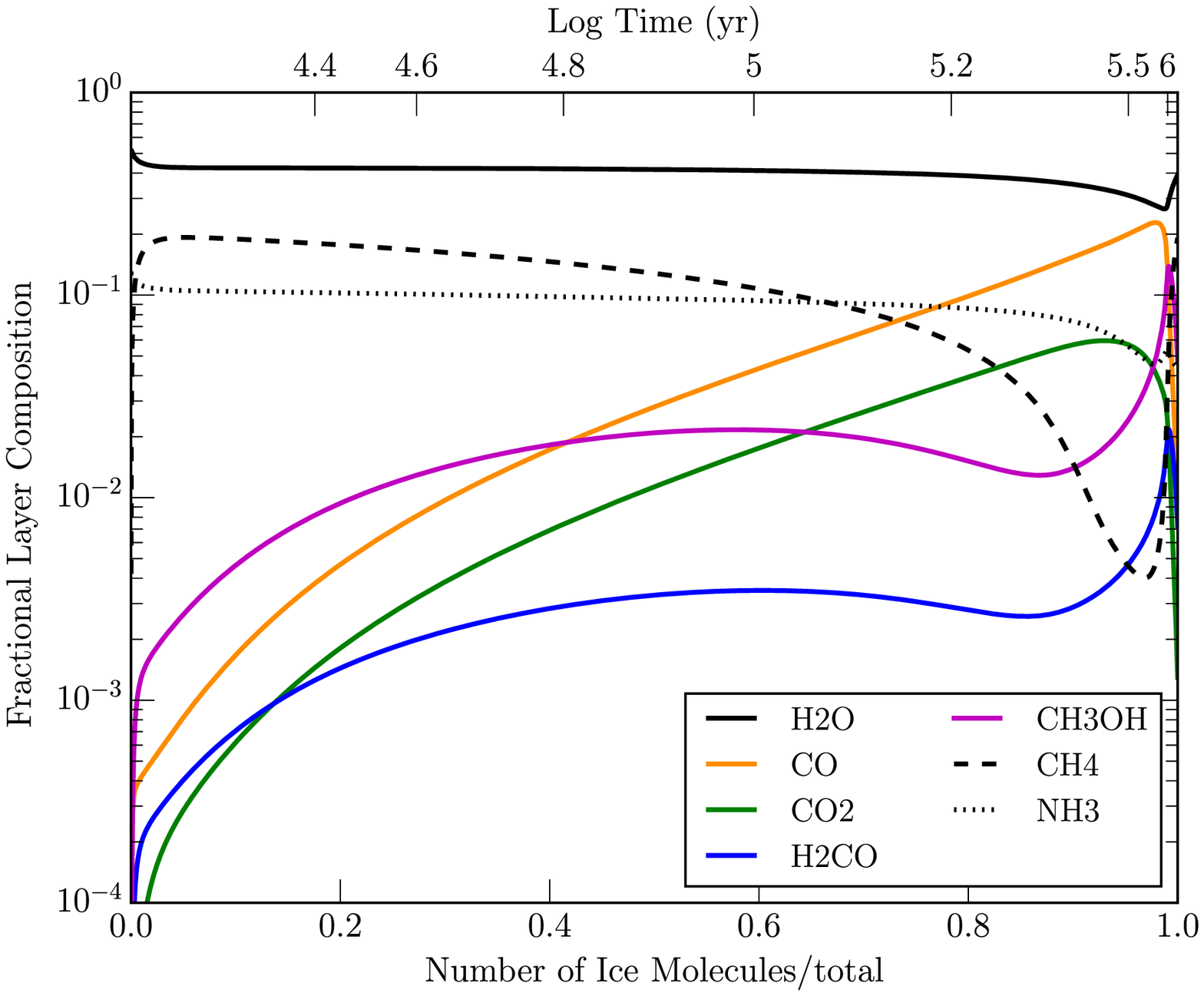}
        }%
        \subfigure[Grain Size 5]{%
            \label{fig:sixth}
            \includegraphics[width=0.46\textwidth]{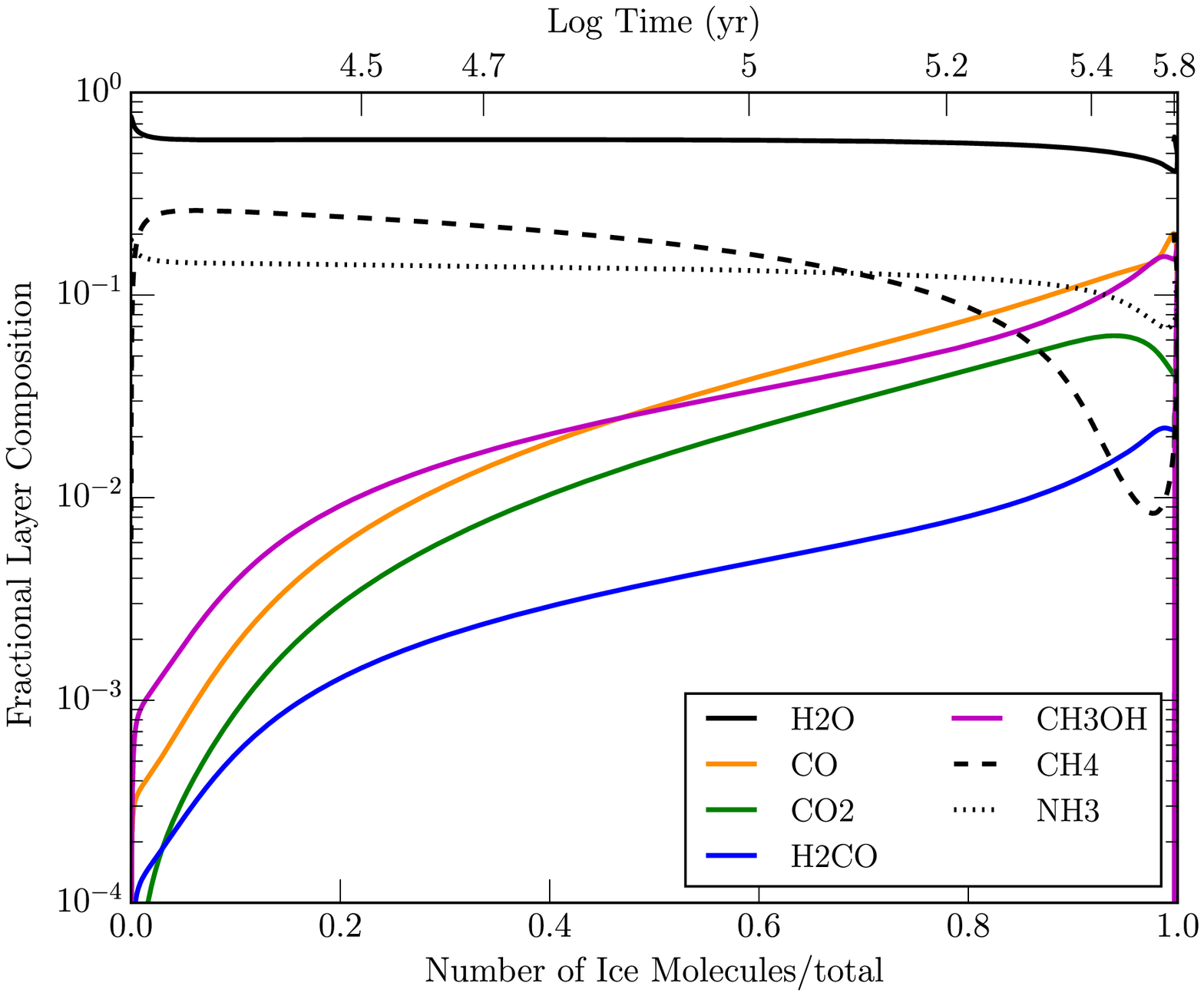}
        }%
    \end{center}
    \caption{%
        Chemical evolution for models 5G\_T8\_DIST (left column) and 5G\_T12\_DIST (right column).  Strong differences are present between all three temperature distribution models. 
     }%
   \label{fig:extradistchemevo}
\end{figure*}

Figure~\ref{fig:extradistchemevo} shows the chemical evolution of the two models. The extreme temperature of grain size 1 in 5G\_T12\_DIST produces a $\mathrm{CO_2}$-dominated mantle, with CO comprising less than $10^{-4}$ of a layer in most of the mantle layers. Due to the majority of the dust cross-sectional area coming from the smallest grain size, the aggregate chemistry of 5G\_T12\_DIST is CO$_2$ dominated. Grain size 5 shows fairly efficient production of $\mathrm{CH_3OH}$, but the low accretion rates caused by small population of the largest grain size causes the effects of grain 5 to be muted in the aggregate mantle. Model 5G\_T8\_DIST has a CO dominated chemistry, with the temperature of only the smallest grain rising above 12 K, and only for the beginning of mantle accumulation. As the temperature of the smallest grain population drops, the chemistry becomes fairly uniform across grain sizes. This suggests that the temperature distribution may be most important when the grain temperature approaches a threshold temperature for an important surface reaction, e.g. the 12 K threshold for efficient formation of CO$_2$ from CO.

\begin{figure*}
     \begin{center}
        \subfigure[Aggregate Ice Mantle]{%
            \label{fig:5gcollgr0}
            \includegraphics[width=0.46\textwidth]{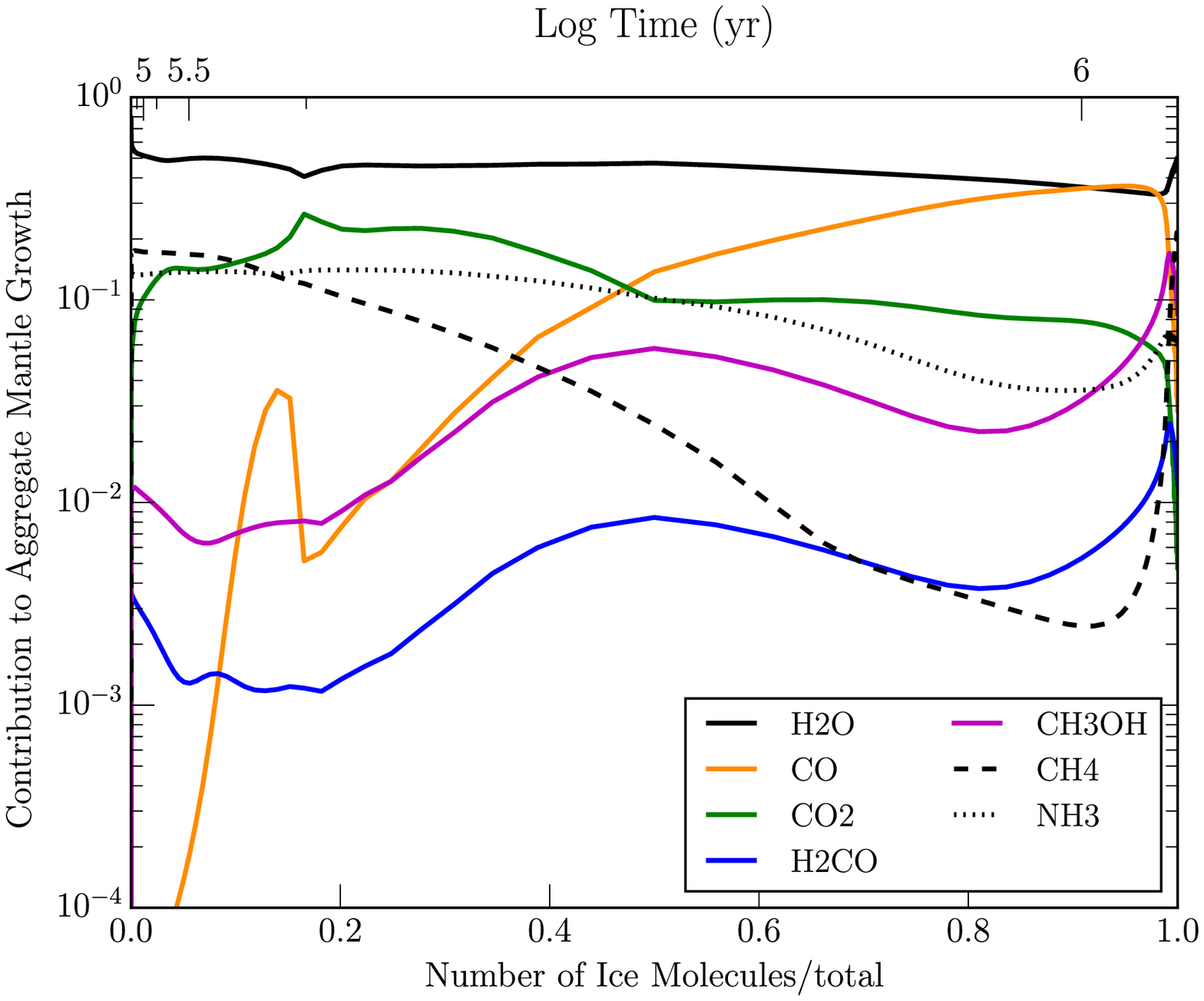}
        }%
        \subfigure[Grain Size 1]{%
           \label{fig:5gcollgr1}
           \includegraphics[width=0.46\textwidth]{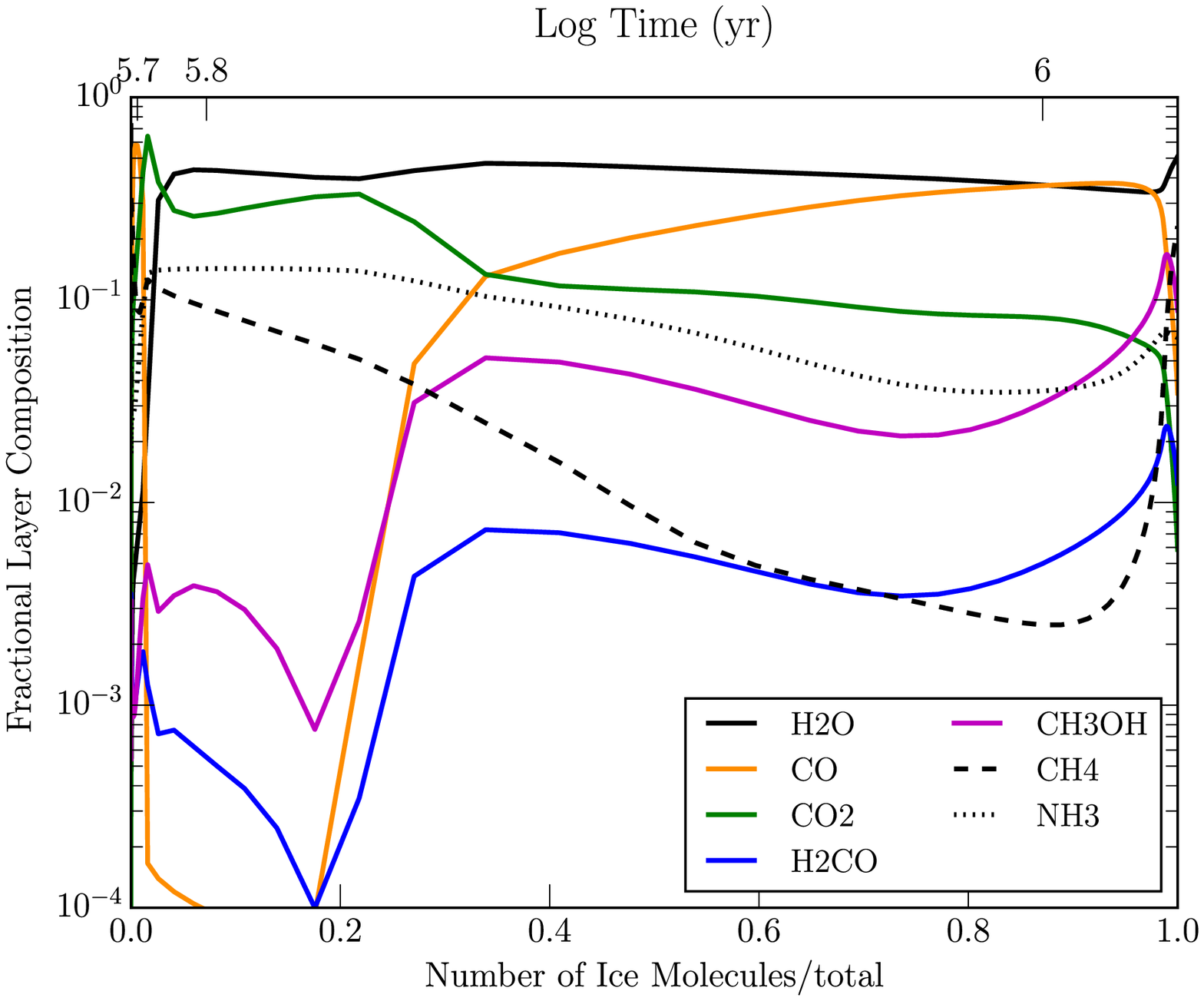}
        }\\ 
        \subfigure[Grain Size 2]{%
            \label{fig:5gcollgr2}
            \includegraphics[width=0.46\textwidth]{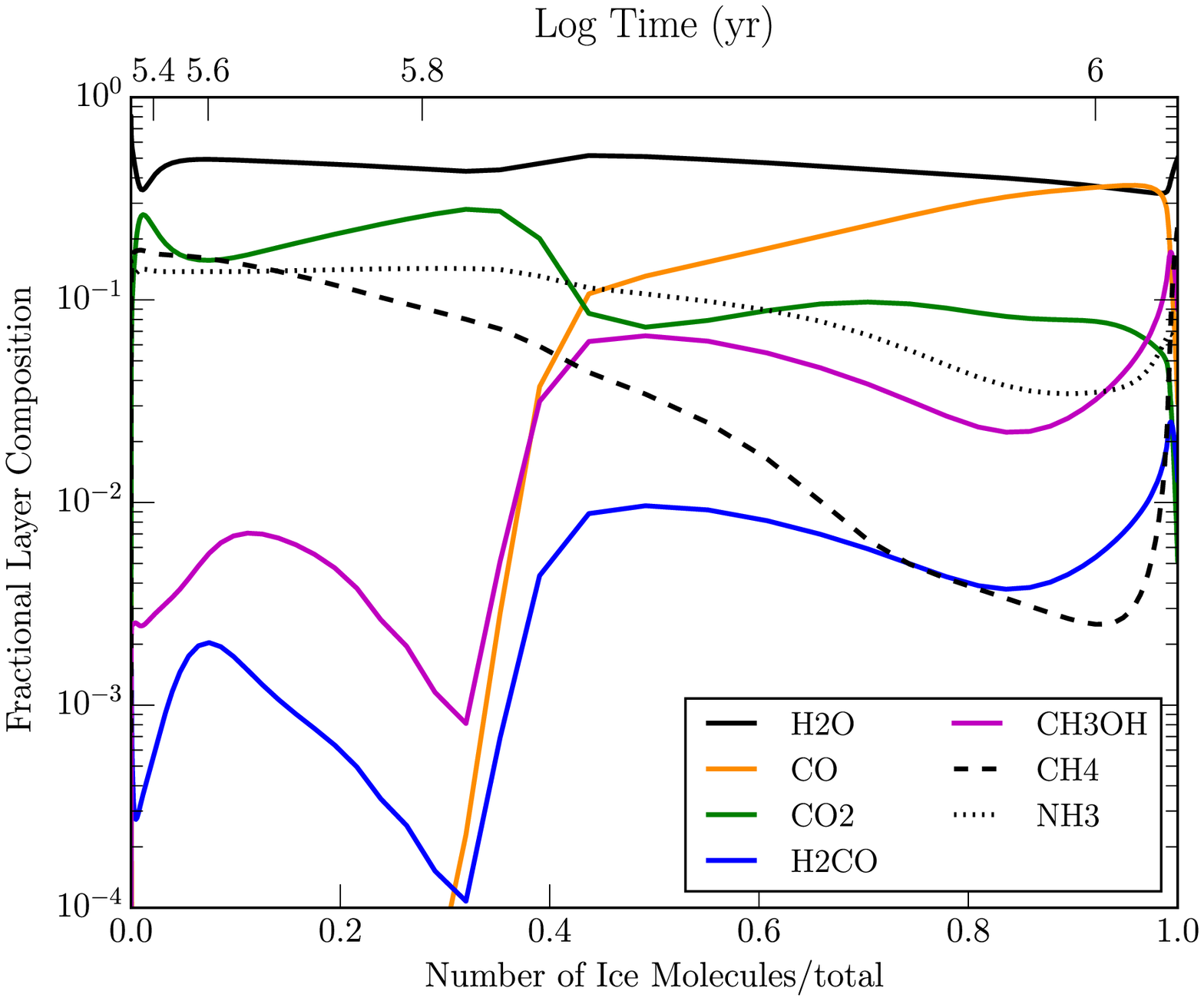}
        }%
        \subfigure[Grain Size 3]{%
           \label{fig:5gcollgr3}
           \includegraphics[width=0.46\textwidth]{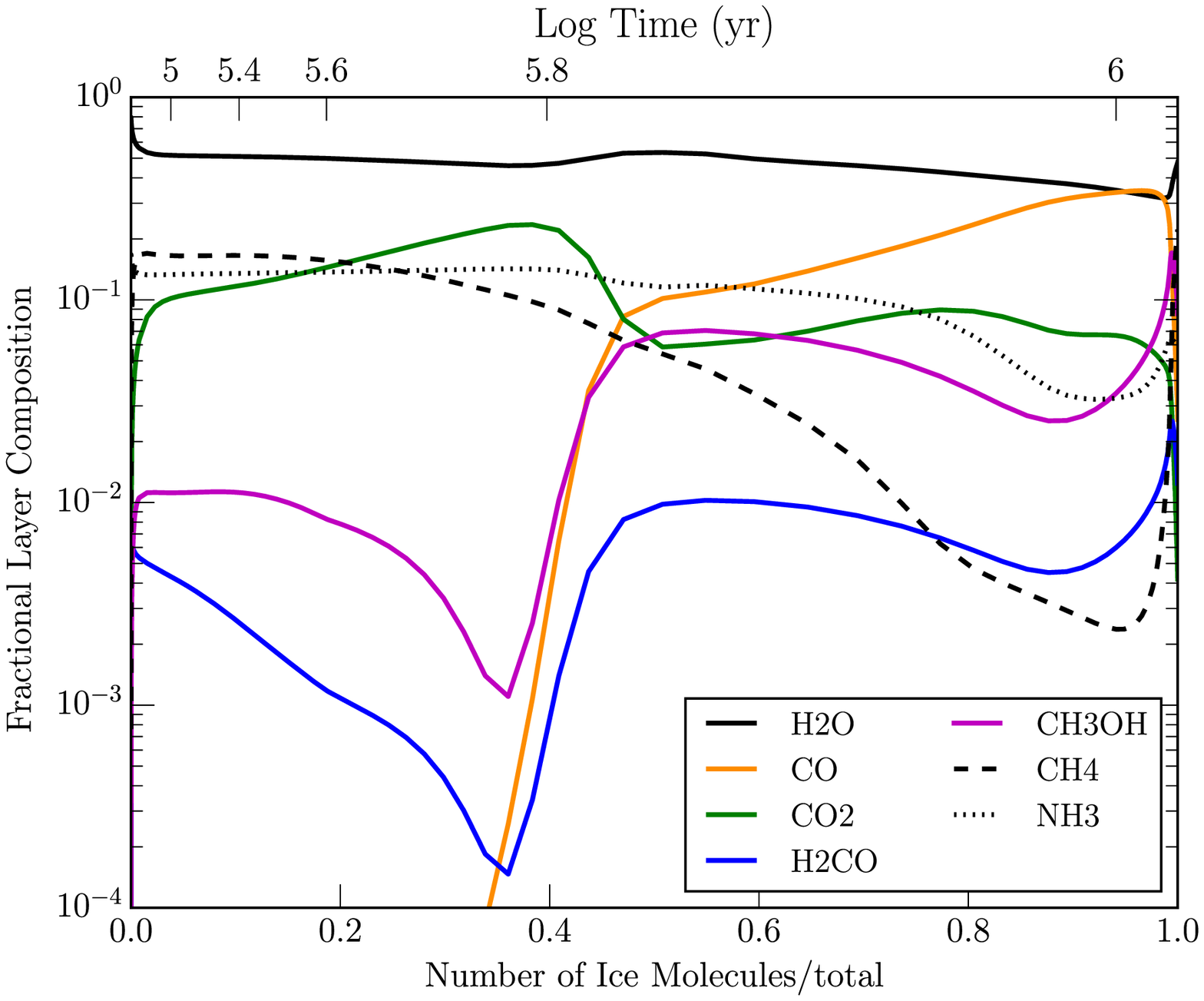}
        }\\ 
        \subfigure[Grain Size 4]{%
            \label{fig:5gcollgr4}
            \includegraphics[width=0.46\textwidth]{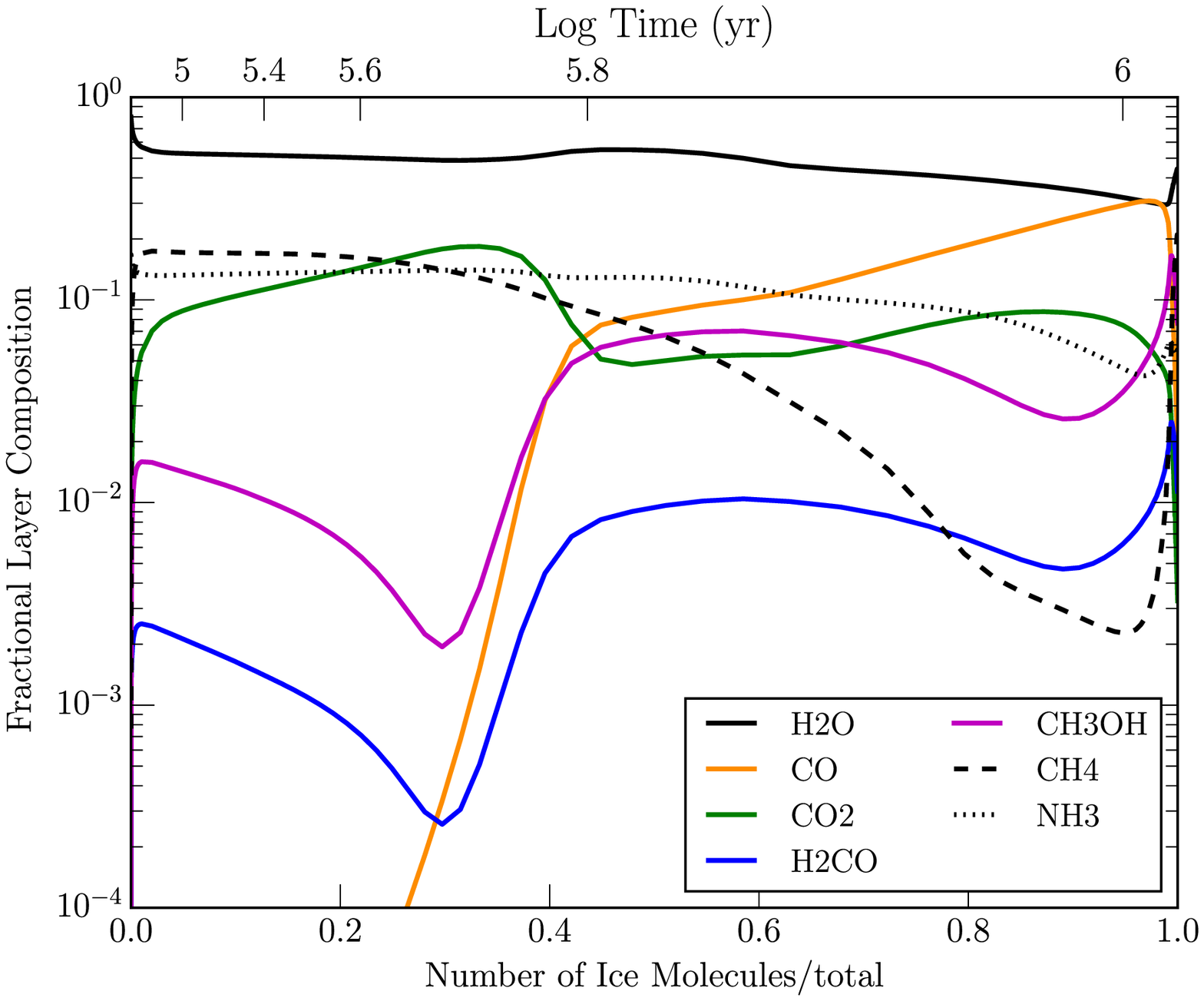}
        }%
        \subfigure[Grain Size 5]{%
            \label{fig:5gcollgr5}
            \includegraphics[width=0.46\textwidth]{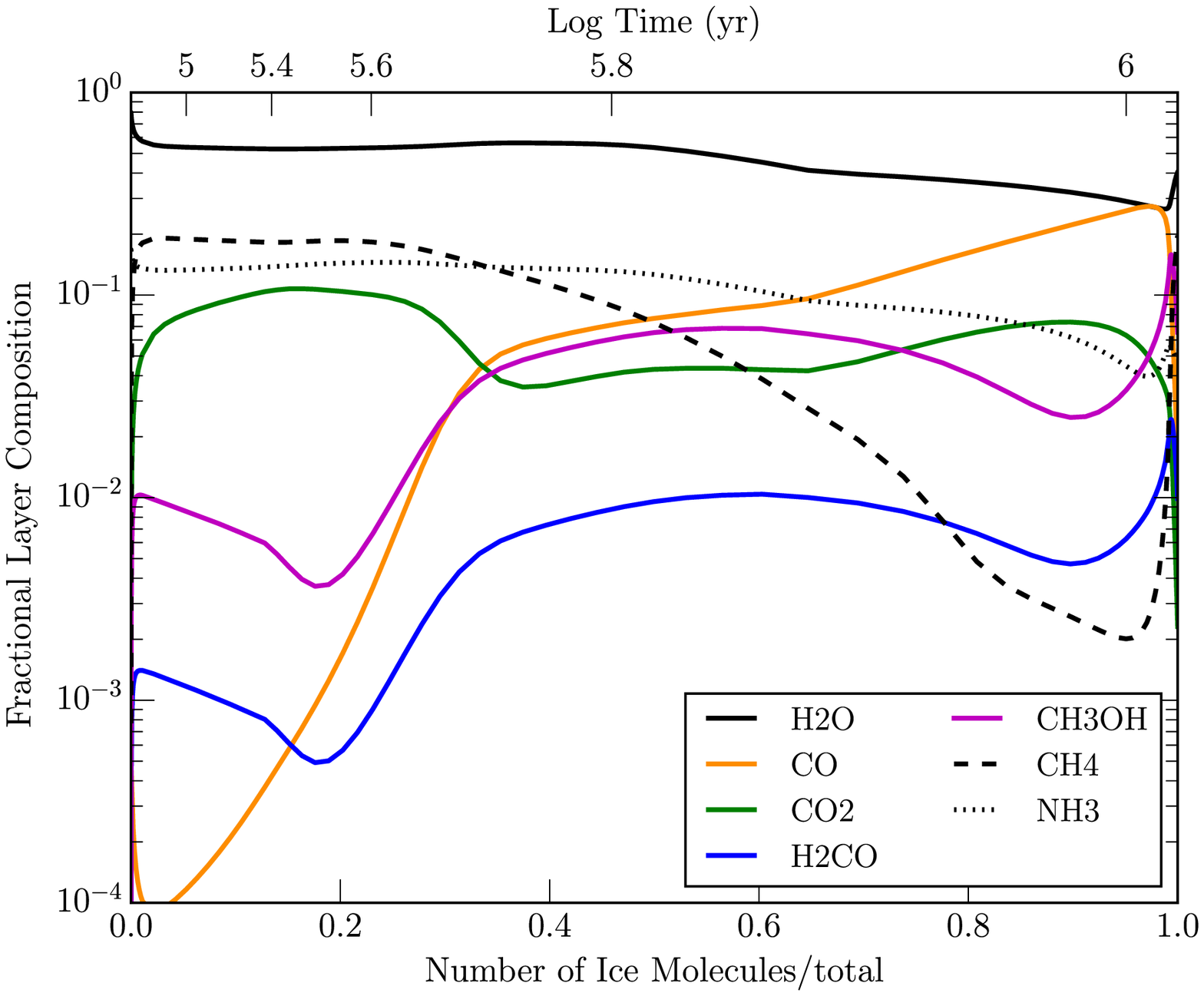}
        }%
    \end{center}
    \caption{%
        Chemical evolution for model 5G\_COLL. The upper left panel shows the evolution of the aggregate ice mantle, while the other panels show the evolution for the smallest grain (1) to the largest grain (5).
     }%
   \label{fig:5gcoll}
\end{figure*}

\begin{figure*}
    \begin{center}
      \includegraphics[width=0.46\textwidth]{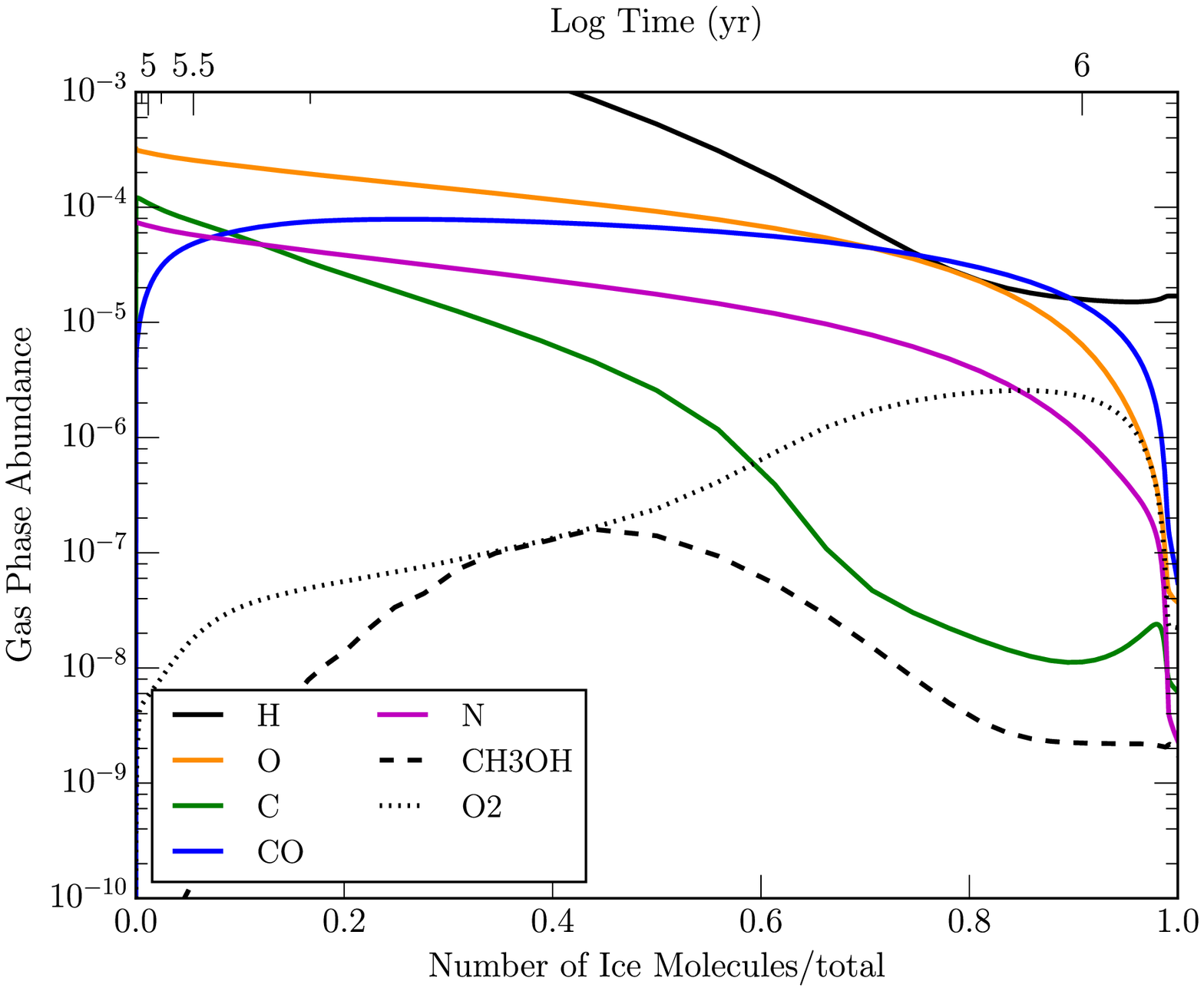}
    \end{center}
    \caption{%
        Gas-phase chemical evolution for model 5G\_COLL, with abundances relative to $\mathrm{n_H}$.
     }%
   \label{fig:5gcollgas}
\end{figure*}

\subsection{Chemical Evolution - Collapse Model}

Figure~\ref{fig:2e4Collgrowth} shows the grain size and temperature evolution for model 5G\_COLL. The dust temperature for this model is determined by the power law $T_d \propto a^{-1/6}$, which is normalized to the temperature of a 0.1 $\mathrm{\mu m}$ grain, as determined by the visual extinction at any moment in the model run \citep{Garrod11}. The time of highest rate of collapse in the model can be seen in the figure as the time at which most accretion occurs, around $\mathrm{6 \times 10^5 yr}$. Prior to the main collapse phase, the small grains are much hotter than in any of the static-cloud models. Post-collapse, the grain temperatures are comparable to the final temperatures for model 5G\_T8\_DIST, see Figure~\ref{fig:extradistchemevo}.

The chemical evolution of the collapse-model grain mantles is shown in Figure~\ref{fig:5gcoll}. Cusp-like features are present in the aggregate mantle plot due to the rapid changes in surface chemistry during the collapse. Prior to collapse, the dust temperature of the smallest grains is greater than 20 K. At this temperature, accreted atomic hydrogen and oxygen quickly desorb, and the time spent on the grain surface is not sufficient to produce OH efficiently. During this stage, the surface is predominantly covered with CO. However, once the small grains drop below 19 K, atomic oxygen remains on the surface long enough to react with atomic hydrogen to form OH, and OH concentration jumps by eight orders of magnitude. Surface CO is then efficiently converted to CO$_2$ via CO + OH.  This transition in behavior can be seen in Panel~\ref{fig:5gcollgr0} where CO shows a peak in fractional layer abundance near 15\% of total ice deposited. Past the peak, the contribution to aggregate CO fraction from grain one becomes negligible, as its CO is efficiently converted into CO$_2$. After the majority of the collapse has taken place and the temperature drops further, the smallest grain drops below 12 K, the temperature required for efficient CO$_2$ production.  The aggregate mantle returns to a CO$_2$/CO ratio \textless 1.

The evolution of selected gas-phase species is shown in Figure~\ref{fig:5gcollgas}. The gas-phase evolution has not changed considerably from the models of \citet{Garrod11}. The largest difference occurs in methanol, with elevated abundances in the current 5G\_COLL model. Methanol is produced on grains surfaces, but a fraction (\textless 1 \%) is returned to the gas phase through the reactive-desorption mechanism \citep{Garrod07}. Gas-phase methanol abundance is highly sensitive to the barriers set for hydrogen abstraction from surface methanol by atomic H. The more rapid abstraction produced in this model results in more CH$_3$O/CH$_2$OH, which quickly recombines with H, resulting in more methanol desorption. The methanol desorbed into the gas phase is nevertheless only a small fraction of the total produced on the grains.

Compared to 5G\_T10\_DIST, the general composition of the aggregate mantle has been enriched in CO and depleted in $\mathrm{H_2O}$ and $\mathrm{CH_4}$. CO$_2$ abundance is comparable in both models, indicating the mean dust temperature during collapse is somewhat similar to the DIST model. Gas-phase abundances of grain-surface products closely follow their production due to desorption of surface species. Gas-phase CO$_2$ peak abundances are comparable in 5G\_COLL and 5G\_T10\_DIST models, though final collapse values are greatly reduced due to a decrease in surface abundance, coincident with the decrease in grain temperature. 

Methanol formation in the collapse model is fairly stable, showing the largest variation when a grain crosses from CO$_2$-dominated to CO-dominated surface composition. With abundant surface CO, the hydrogenation chain to $\mathrm{CH_3OH}$ occurs more frequently. The grains in the collapse model spend the majority of time under 12 K, which leads to comparable methanol abundances to the UNIF models with T $\lesssim$ 12 K.

\begin{deluxetable}{lrrrrrrr}
\tablecolumns{7}
\tablewidth{0pc}
\tablecaption{Gas-to-dust ratios, calculated as the number of hydrogen atoms to the number of grains, for classes of models with a given number of grain sizes. Each column presents values from a single model, with the gas-to-dust ratio for a given size category read from the rows, i.e. the smallest grain in each model is the first row and the largest grain in the last non-empty row.}
\tablehead{
\colhead{$\mathrm{N_{b}}$} & \colhead{$\mathrm{GD_{gr1}}$}   & \colhead{$\mathrm{GD_{gr2}}$}    & \colhead{$\mathrm{GD_{gr3}}$} &
\colhead{$\mathrm{GD_{gr4}}$}    & \colhead{$\mathrm{GD_{gr5}}$}   & \colhead{$\mathrm{GD_{gr6}}$} & \colhead{$\mathrm{GD_{gr7}}$}}
\startdata
1      & $5.694\times 10^9$  & $5.737\times 10^9$ & $5.921\times 10^9$ & $6.234\times 10^9$ & $6.632\times 10^9$ & $7.081\times 10^9$ & $7.564\times 10^9$ \\
2      & -  & $7.628\times 10^{11}$ & $1.542\times 10^{11}$ & $7.189\times 10^{10}$ & $4.689\times 10^{10}$ & $3.614\times 10^{10}$ & $3.059\times 10^{10}$ \\
3      & -  & - & $4.018\times 10^{12}$ & $8.289\times 10^{11}$ & $3.316\times 10^{11}$ & $1.845\times 10^{11}$ & $1.237\times 10^{11}$ \\
4      & -  & - & - & $9.558\times 10^{12}$ & $2.345\times 10^{12}$ & $9.415\times 10^{11}$ & $5.001\times 10^{11}$ \\
5      & -  & - & - & - & $1.658\times 10^{13}$ & $4.805\times 10^{12}$ & $2.022\times 10^{12}$ \\
6      & -  & - & - & - & - & $2.453\times 10^{13}$ & $8.178\times 10^{12}$ \\
7      & -  & - & - & - & - & - & $3.307\times 10^{13}$ \\
\enddata
\label{table:gdvalues}
\end{deluxetable}

\subsection{Effect of Discretization and Number of Grain Sizes}

To test the effects of discretizing the grain size distribution and to determine if five grain sizes are sufficient to reproduce the effects of a continuous distribution, we use models with various numbers of grain sizes.  The number of grain sizes is used to bin the power law from Equation \ref{eq:powerlaw}, with calculated grain abundances given in Table \ref{table:gdvalues}. Shown in Table~\ref{table:numgrs} are the results of collapse models ranging from two grains to eleven.  Figure 9 shows the aggregate mantles of the models for comparison.  The effects of discretizing the grain size distribution are seen in the abundance discrepancies for the two and three grain models. Note that the number of peak-like deviations of CO at early time correspond to the number of grain sizes that start at a dust temperature too high for atomic oxygen to form OH on the surface. Beyond five grains, the differences in species abundances drop to minimal levels.  This supports a choice of four or five grains to capture the complexity of the size distribution while minimizing computational runtime.

\begin{deluxetable}{c c c c c c c c}
\tablecolumns{6}
\tablewidth{0pc}
\tablecaption{Mantle abundances at $\mathrm{5 \times 10^6}$ years for models with varying numbers of grain sizes. The value of $\mathrm{H_2O}$ listed is with respect to total atomic hydrogen abundance, while the values for other species are given with respect to the $\mathrm{H_2O}$ abundance.}
\tablehead{
\colhead{Model} & \colhead{$\mathrm{H_2O}$}   & \colhead{$\mathrm{CO}$}    & \colhead{$\mathrm{CO_2}$}   &
\colhead{$\mathrm{CH_3OH}$}  & \colhead{$\mathrm{H_2CO}$}  & \colhead{$\mathrm{CH_4}$}  & \colhead{$\mathrm{NH_3}$}}
\startdata
 2G\_COLL    & 1.485(-4) & 35.4 & 30.6 & 7.29 & 1.10 & 11.1 & 21.6 \\ 
 3G\_COLL    & 1.508(-4) & 34.9 & 29.4 & 7.37 & 1.13 & 11.5 & 21.5 \\
 4G\_COLL    & 1.513(-4) & 33.9 & 29.7 & 7.32 & 1.12 & 12.0 & 21.7 \\
 5G\_COLL    & 1.507(-4) & 34.1 & 30.0 & 7.33 & 1.12 & 11.9 & 21.8 \\
 6G\_COLL    & 1.506(-4) & 34.2 & 30.0 & 7.37 & 1.13 & 11.8 & 21.7 \\
 7G\_COLL    & 1.507(-4) & 33.9 & 30.2 & 7.35 & 1.13 & 12.0 & 21.8 \\
 11G\_COLL    & 1.460(-4) & 34.1 & 31.0 & 7.32 & 1.13 & 11.8 & 21.8 \\
\enddata
\label{table:numgrs}
\end{deluxetable}

\begin{figure*}
     \begin{center}
        \subfigure[2 Grains]{%
            \label{fig:first}
            \includegraphics[width=0.46\textwidth]{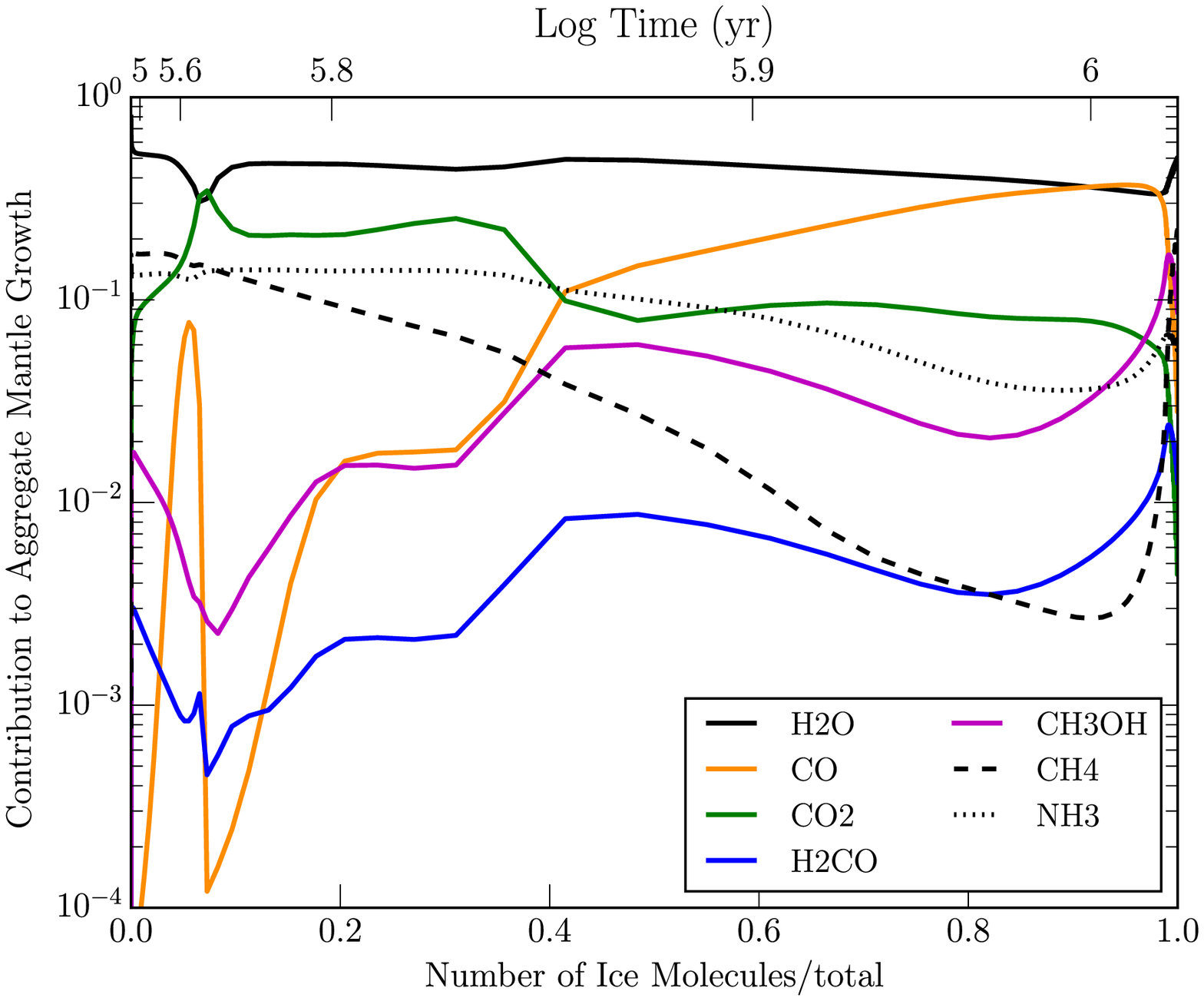}
        }%
        \subfigure[3 Grains]{%
           \label{fig:second}
           \includegraphics[width=0.46\textwidth]{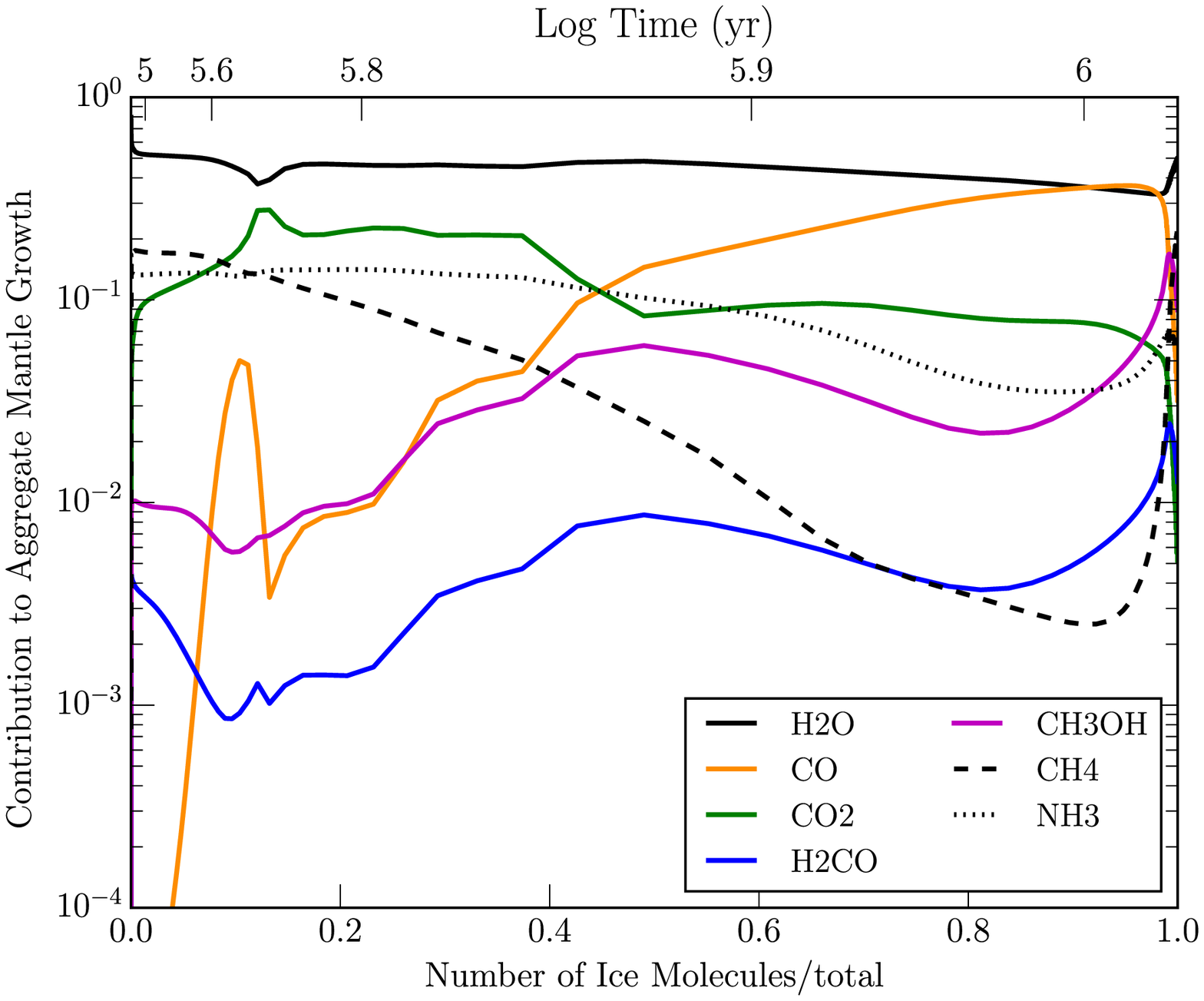}
        }\\ 
        \subfigure[4 Grains]{%
            \label{fig:third}
            \includegraphics[width=0.46\textwidth]{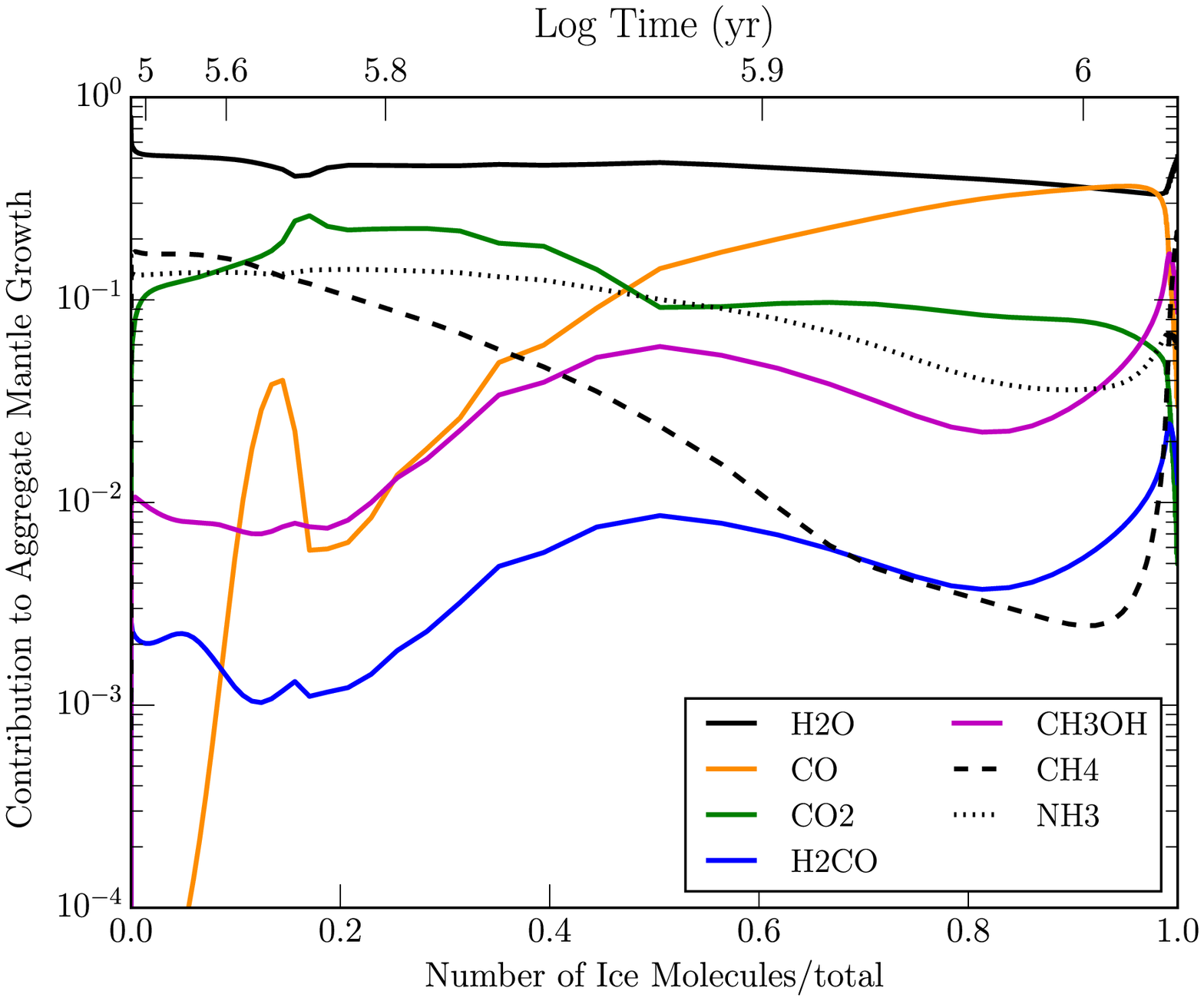}
        }%
        \subfigure[6 Grains]{%
            \label{fig:fourth}
            \includegraphics[width=0.46\textwidth]{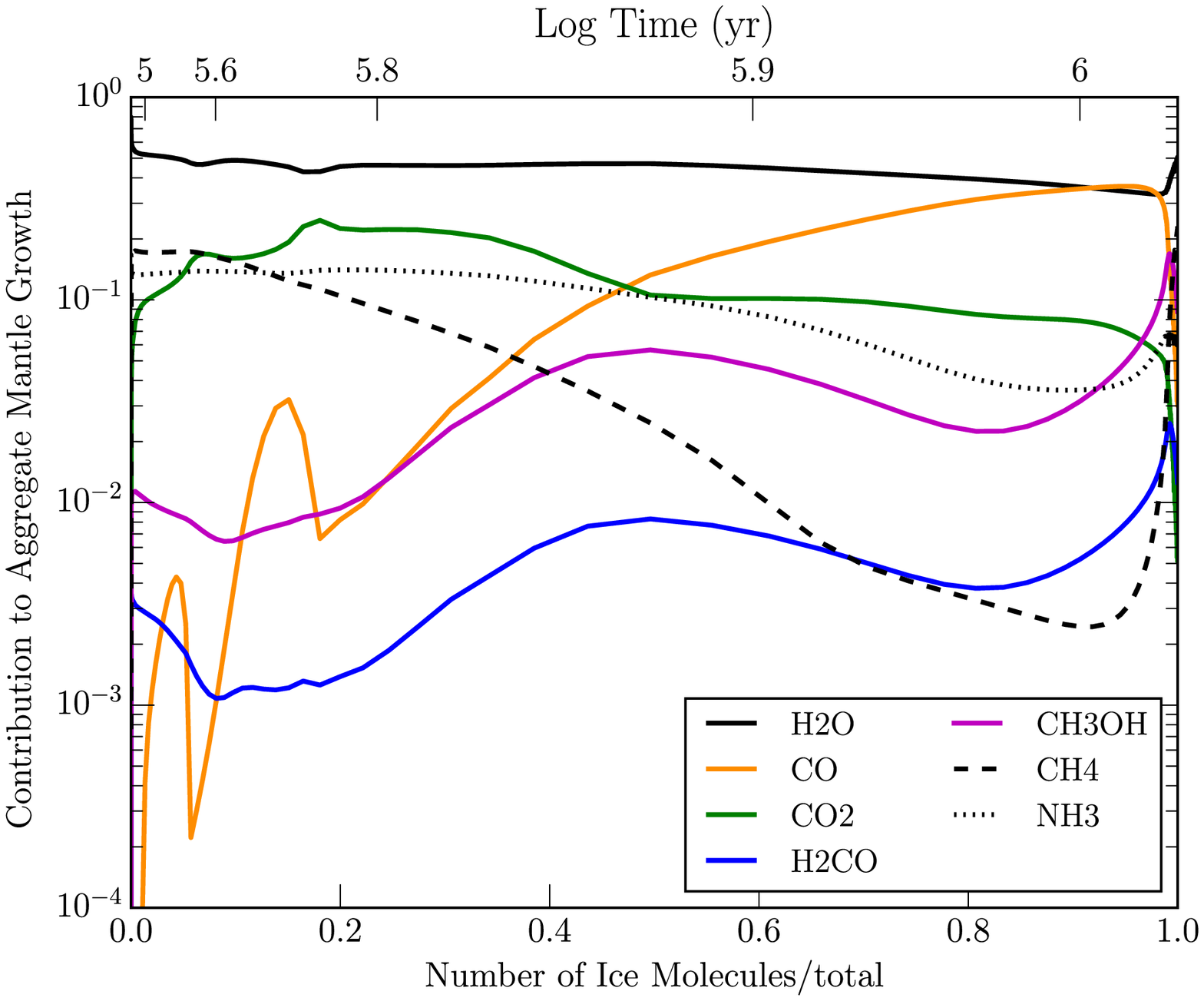}
        }\\ 
        \subfigure[7 Grains]{%
            \label{fig:fifth}
            \includegraphics[width=0.46\textwidth]{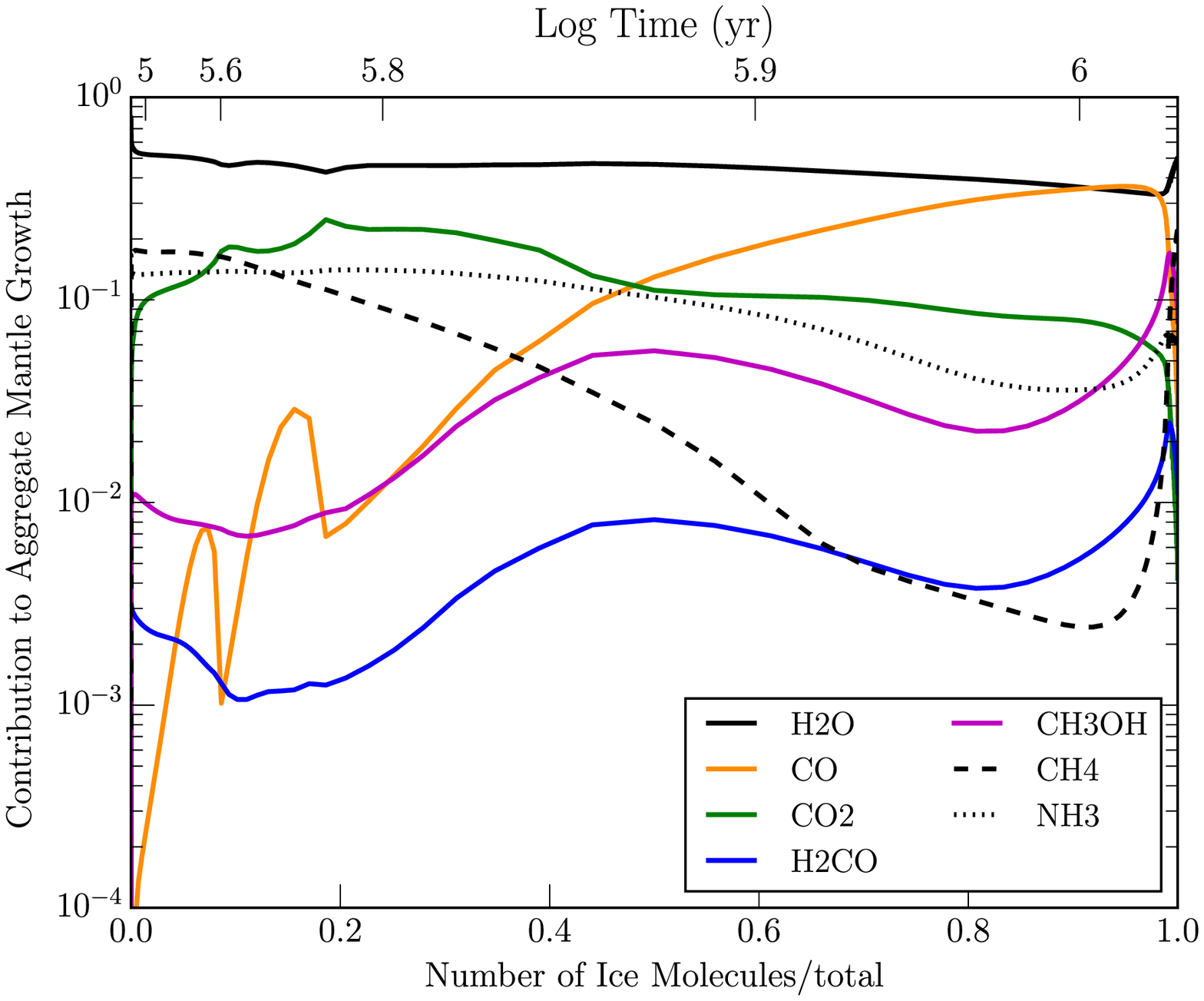}
        }%
        \subfigure[11 Grains]{%
            \label{fig:sixth}
            \includegraphics[width=0.46\textwidth]{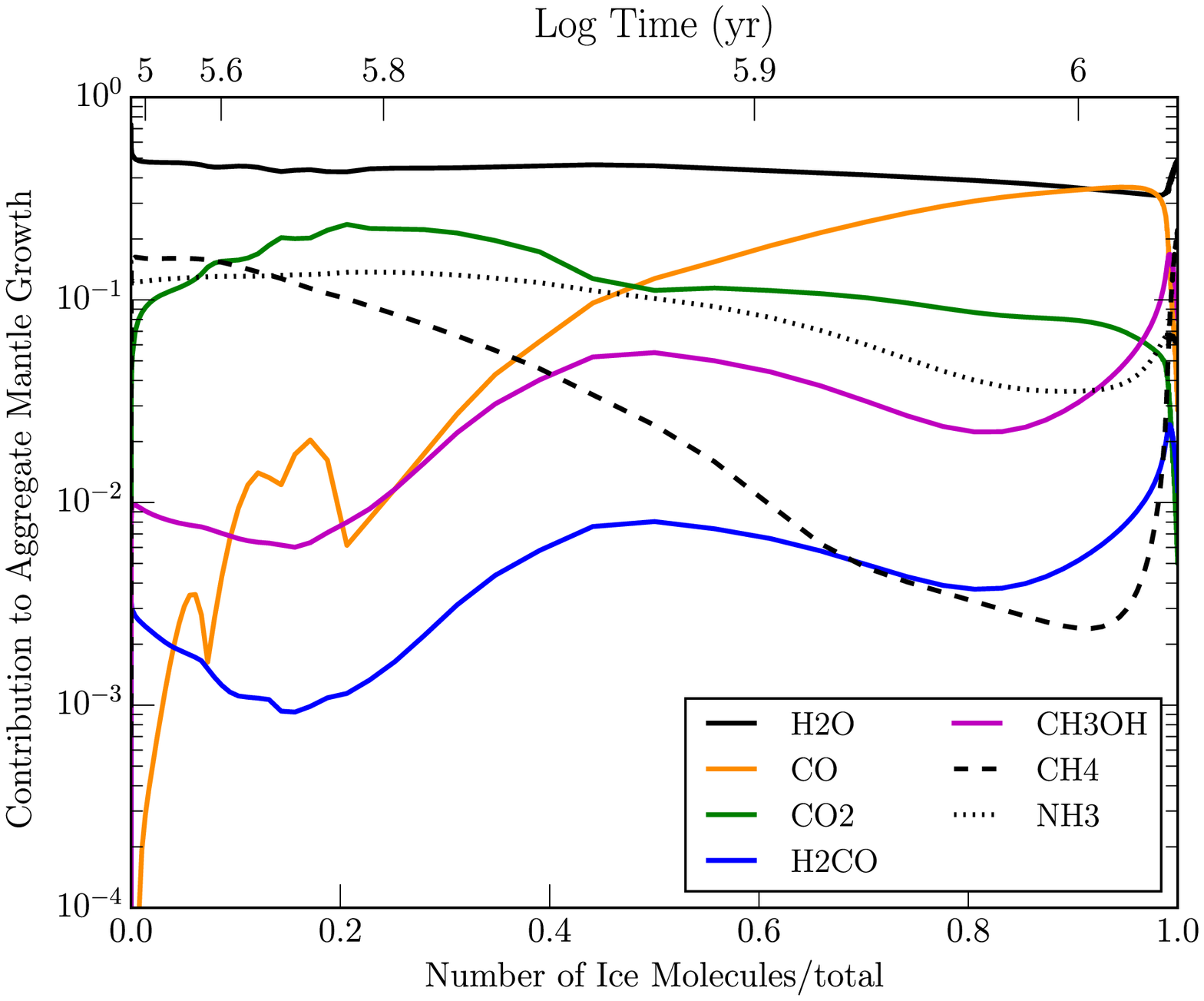}
        }%
    \end{center}
    \caption{%
        Aggregate mantles for collapse models with varying numbers of grain sizes.
     }%
    \label{fig:subfigures}
\end{figure*}

\section{Discussion}

The behavior of models with uniform temperature across a grain-size distribution shows surprising similarity to the single grain counterparts, indicating that simply introducing the distribution without any associated change in temperature has little effect on the chemistry. 

A fraction of surface reactions are well-modeled with rate equations, while the rest are altered by stochastic effects. The cross-sectional area of a dust grain is used to calculate the accretion rates for gas-phase species; if the surface population of all reactants drops to of order unity, the modified rate equations are used. These modifications were the primary mechanism expected to alter the surface chemistry for the class of models with uniform temperature.

The models find the same rate treatment is appropriate for a given reaction on all grain sizes, for nearly all ($\sim$90\%) surface reactions. There are quantitative differences in ice composition between the grain sizes in the UNIF models; however, these differences follow a relationship that is reasonably reproduced by a single grain size model. This is shown in Table \ref{table:mantlefraction} by comparing single grain models to UNIF models with equal temperature. Grain-size-dependent behavior is observed in the UNIF models as an increase in the abundance of the hydrogenated species $\mathrm{CH_4}$ and $\mathrm{NH_3}$ with increasing grain size. This occurs due to the stochastic nature of the hydrogenation reactions; the intermediate steps from atomic carbon and nitrogen to their hydrogenated forms are rate-limited by the accretion of hydrogen, which is larger for grains with a larger cross section.

The inclusion of a temperature distribution across grain sizes creates a more varied composition across the grain sizes.  The conversion of CO to CO$_2$ occurs via an efficient pathway above a threshold temperature of $\sim12\mathrm{K}$. CO is strongly depleted for grains with a temperature above this threshold, and CO$_2$ is the primary carbon-bearing surface species. Correlation of $\mathrm{H_2CO}$ and $\mathrm{CH_3OH}$ abundances to CO abundance follow from their formation via CO hydrogenation; the low abundance of CO on small, warm grains causes a depletion in methanol on those grains. It should be noted that in absolute quantity, the smallest grains still possess the most methanol, due to their greater total number and summed cross section; rather, it is a depletion in the amount of methanol ice when expressed as a fractional abundance with respect to water ice.

The collapse models explore higher temperatures at early times on the smallest grains, shown in Figure \ref{fig:2e4Collgrowth}. The accretion closely follows the collapse, occurring as a burst rather than a sustained rate. The aggregate mantle plots have cusp-like behavior due to grain temperatures rapidly decreasing during collapse, with the most pronounced compositional changes on the smallest grains. At early times the smallest grains are sufficiently warm that atomic species do not have time to react before desorbing. As the collapse proceeds, accretion rates increase and the dust temperature drops. Once a dust temperature of ~20 K is reached, $\mathrm{O} + \mathrm{H} \rightarrow \mathrm{OH}$ becomes competitive as a sink for atomic oxygen, and the binding energy of OH is large enough to secure it on the grain surface. Through reaction with OH, surface CO abundance sharply plummets to a value of order unity; the less-efficient modified rate prevents CO + OH from further reaction. This is seen in the aggregate mantle plot \ref{fig:5gcollgr0} as a sharp dip in fractional layer abundance of CO. CO$_2$ is efficiently formed until the dust temperature crosses the 12 K threshold.The initial temperature of the largest grains in the 5G\_COLL model is sufficiently low that atomic species readily stick to the surface.  The elevated abundance of atomic hydrogen creates an enrichment of hydrogenated species - $\mathrm{H_2O}$, $\mathrm{CH_4}$, and $\mathrm{NH_3}$.

The set of collapse models with varying numbers of grain sizes points to the threshold temperature detailed above. For less than five grain sizes, only the smallest grain size in a given model is too hot for atomic species to react. For six to ten grain sizes, there are two, while the eleven grain model shows three grains with a thin surface layer of accreted CO at early time. Examining the aggregate mantle abundances, all models agree to ~5\% accuracy for most major species.  Five grains was chosen for models as a compromise between optimal sampling of the grain size distribution and computational complexity.

When comparing the grain size distribution models to those with a single grain size, the primary effects we expect to see should be evident of a difference in chemistry on the small, hot grains versus the large, cool grains. These effects are muted in the UNIF and DIST models because the small grains dominate the distribution of cross-sectional area; in the 5G\_T10\_DIST model, nearly 60\% of total grain species are found on the smallest grain, and 80\% on the two smallest grains. However, the difference is more pronounced in the 5G\_COLL model. With grain temperatures above 20 K, the two smallest grains fail to retain significant surface species until collapse. During this time, the third grain size has the most grain species, at around 42\%. Post-collapse, the two smallest grains grow substantially and hold roughly 70\% of grain species.

The high-temperature desorptive effect serves to hinder observational differentiation between single-grain and multi-grain models; the change in chemistry we might expect to see on small grains is absent, because their surface does not allow for adsorbed species to efficiently react.  The dominant grain mantle is the warmest surface under 20 K, which will preferentially form CO$_2$ ice over CO ice if that surface is above 12 K. Observational constraints may be more forthcoming when considering grains in a post-collapse warm-up phase associated with hot cores in the star formation process, where more refractory species are found initially on grain surfaces.  Even though in the denser conditions of a hot core, it is likely the dust and gas temperatures would be well coupled, flattening out the temperature distribution across the grain sizes, the enriched state of the smaller grains could be enhanced by the thin nature of ice mantles on the smallest grains; the top layer of the smallest grains in the 5G\_COLL model contains 5.1\% of the total ice, and 43.4\% of the ice is found in the uppermost ten layers. A significant fraction of the mantle could then be involved in surface chemistry such as diffusion, reaction, and evaporation. It is worth noting that the heating of small grains can be stochastic, and the time variation of small grain temperatures can be extreme for the smallest grains \citep[see][]{Cuppen06}.

We therefore suggest that, under diffuse to translucent conditions where grain mantles are beginning to form, the smallest grains should be essentially bare, with substantial proportions of the ice abundance stored on the mid-size to large grains. This balance should shift during cloud collapse so that small grains hold the majority of ice material, albeit in thinner ice mantles than are formed on the larger grains.

Past work \citep{Garrod11} found a similar temperature threshold for CO$_2$ production; the model results were interpreted as an explanation for the abundance of polar and apolar CO and CO$_2$ ices at varying temperatures in dark clouds \citep[for observations, see][]{Whittet09,Cook11}. This interpretation is not changed by the results of these grain-size distribution models. The effect of forming CO$_2$-rich, CO-poor mantle layers at high temperature is still present, as is the CO/CO$_2$ abundance ratio of 2-4 at temperatures lower than 12 K. It should be noted that the collapse models end at a density equivalent to that found in dark clouds (n$_\mathrm{H}$ = 4$\times$10$^4$ cm$^{\mathrm{-3}}$), by which point gas-phase CO is not fully depleted. In some sources, significantly greater gas densities may be achieved, with correspondingly higher CO depletions.

The most important attributes (assumptions) of the distribution are how the cross-sectional area is distributed across grain sizes and how the temperature of grains is determined. For the Mathis power law, the $area$ is distributed as  $\frac{dX_{tot}}{da} \propto a^{-1.5}$, placing the majority of accretion onto small grains. This smoothes the power law into a flatter effective distribution as small grains accrete, shown in the initial and final radii in Table \ref{table:graingrowth}. Future work will consider more complex grain distributions, like those of \citet{Weingartner01}. Distributions that place the majority of grain surface area on large grains could affect the aggregate chemistry. The temperature distribution assumes that the grains are heated by an ambient radiation field and that the dust absorption efficiency is unity across all grains, allowing a simple relationship between the temperatures of each grain size to be assigned. Future modeling will take into account explicitly the optical properties of each grain size, although a full treatment of stochastic heating of the smallest grains is unlikely to be achievable using a rate-based gas-grain chemical model.

\section{Conclusions}

We summarize our main conclusions from this study:
\begin{itemize}
\item Inclusion of a grain size distribution with uniform grain temperature does not strongly affect the model at dark cloud temperatures and densities, despite the inclusion of modified rates. This reinforces the results of Acharyya (2011).
\item Consideration of a grain temperature distribution causes variation in ice production between grain sizes, with CO$_2$ ice favored over CO above 12 K. The small grains are warmest and most abundant, leading to an increase in CO$_2$ ice when comparing models with a dust-temperature distribution against models that use a uniform dust temperature.
\item The collapse model shows unique behavior due to low initial $\mathrm{A_v}$ and high initial temperatures on the smallest grains. Above 20 K, accreted atomic species desorb before reacting with other surface species, with only heavy molecular species like CO bound to the surface. As the collapse proceeds, grain temperatures decrease as density increases and ice mantles grow. This leads to a transitory period of CO$_2$ enrichment, similar to models of Garrod \& Pauly (2011).
\item The majority of grain-surface ice material resides on the smallest grain populations, under low temperature/high extinction conditions, due to the greater size of this population. Under more diffuse conditions where the small-grain temperatures exceed 20 K, the main carrier of the ice mantles, such as they are, will be the medium-sized grains.
\item  The collapse models produce an ice mantle on the dust grains, with a varying thickness from 40 to 100 monolayers, from the smallest to largest grains. On the smallest grains the mantles are thin, and 43\% of the total ice is present in the uppermost ten layers; these species are more readily available for surface chemistry, leading to a possible enrichment in chemical complexity during a warm-up phase following the collapse.
\item The number of grain sizes chosen in the discretization of the size distribution shows only a small chemical effect.  The representative grain radius may need to be reduced from 0.1 $\mathrm{\mu m}$ to 1G model values to better represent the Mathis (1977) distribution.  Observable effects may be more pronounced if small grains are allowed an initial reservoir of refractory molecular species in a core warm-up model.
\item The inclusion of a grain-size distribution is a general improvement for chemical models with variable temperature and density. The distribution in size and temperature captures the changing environment experienced by chemical species during the collapse to a dark cloud. For static models, the use of a grain size distribution is less critical; the effects of the distribution can be recreated by a single-grain model with decreased grain radius and elevated grain temperature, due to the weighting effects of the grain size distribution assumed here.
\end{itemize}

\acknowledgements

We thank the referee for helpful comments and suggestions. This work was funded by the NASA Astrophysics Theory Program, grant number NNX11AC38G. 

\bibliography{refs}

\end{document}